\documentclass[journal=cmatex,layout=draft]{achemso}
\setcitestyle{square}
\setkeys{acs}{etalmode=truncate,maxauthors=0}
\usepackage[version=4]{mhchem}
\usepackage[hyperindex]{hyperref}
\hypersetup{colorlinks,citecolor=blue,linkcolor=blue,urlcolor=blue}
\usepackage{units}
\usepackage{amsmath}
\usepackage{booktabs}
\usepackage{etoolbox}
\newcommand{\voc}{V_\text{OC}}
\newcommand{\vocrad}{V_\text{OC,rad}}
\newcommand{\jsc}{J_\text{SC}}
\newcommand{\jzero}{J_\text{0}}
\newcommand{\eqepv}{\text{EQE}_\text{PV}}
\newcommand{\eqeel}{\text{EQE}_\text{EL}}
\newcommand{\phibb}{\phi_\text{BB}}

\makeatletter
\patchcmd{\acs@contact@details}{E}{*\,E}{}{}
\makeatother

\author{Tanvi Upreti}
\affiliation{Complex Materials and Devices, Department of Physics, Chemistry and Biology~(IFM), Link\"oping University, 581 83 Link\"oping, Sweden}
\alsoaffiliation{Centre for Advanced Materials, Heidelberg University, Im Neuenheimer Feld 225, 69120 Heidelberg, Germany}

\author{Sebastian Wilken}
\affiliation{Complex Materials and Devices, Department of Physics, Chemistry and Biology~(IFM), Link\"oping University, 581 83 Link\"oping, Sweden}
\alsoaffiliation{Physics, Faculty of Science and Engineering, \AA{}bo Akademi University, Porthansgatan 3, 20500 Turku, Finland}

\author{Huotian Zhang}
\affiliation{Biomolecular and Organic Electronics, Department of Physics, Chemistry and Biology~(IFM), Link\"oping University, 581 83 Link\"oping, Sweden}

\author{Martijn Kemerink}
\affiliation{Complex Materials and Devices, Department of Physics, Chemistry and Biology~(IFM), Link\"oping University, 581 83 Link\"oping, Sweden}
\alsoaffiliation{Centre for Advanced Materials, Heidelberg University, Im Neuenheimer Feld 225, 69120 Heidelberg, Germany}
\email{martijn.kemerink@cam.uni-heidelberg.de}

\title{Slow Relaxation of Photogenerated Charge Carriers Boosts Open-Circuit Voltage of Organic Solar Cells}

\begin{document}

\begin{abstract}
Among the parameters determining the efficiency of an organic solar cell, the open-circuit voltage~($\voc$) is the one with most room for improvement. Existing models for the description of~$\voc$ assume that photogenerated charge carriers are thermalized. Here, we demonstrate that quasi-equilibrium concepts cannot fully describe~$\voc$ of disordered organic devices. For two representative donor:acceptor blends it is shown that~$\voc$ is actually 0.1--\unit[0.2]{V} higher than it would be if the system was in thermodynamic equilibrium. Extensive numerical modeling reveals that the excess energy is mainly due to incomplete relaxation in the disorder-broadened density of states. These findings indicate that organic solar cells work as nonequilibrium devices, in which part of the photon excess energy is harvested in the form of an enhanced~$\voc$.
\end{abstract}

Organic photovoltaics~(OPVs) achieve quantum yields\cite{Park2009} and fill factors~(FFs)\cite{Gao2020,Li2019} that are competitive with established technologies such as crystalline Si and GaAs. However, the situation is different with the open-circuit voltage~($\voc$). Relative to the energy of the photons absorbed, $\voc$ is low in OPVs, with the consequence that the overall efficiency lags behind their inorganic counterparts.\cite{Polman2016} This makes understanding the nature of these voltage losses and finding strategies to reduce them essential research problems regarding OPVs.

According to current understanding, $\voc$ is generally limited by the splitting of the quasi-Fermi levels of electrons and holes under illumination.\cite{Wurfel2000} Using the principle of detailed balance, this gives 
\begin{equation}
\voc = \frac{kT}{q} \ln \left(\frac{\jsc}{\jzero} + 1\right),
\label{eq:voc}
\end{equation}
where $k$ is the Boltzmann constant, $T$ the temperature, $q$ the elementary charge, $\jsc$ the short-circuit current and~$\jzero$ the dark saturation current. The parameter~$\jzero$ reflects the current associated with thermal excitation of electrons over the band gap; it thus contains all information about recombination losses, either via radiative or nonradiative pathways.\cite{Tvingstedt2016,Cuevas2014} Using the reciprocity relation by Rau,\cite{Rau2007} $\jzero$ can be calculated from the photovoltaic quantum efficiency~$\eqepv$ and the electroluminescence~(EL) quantum yield~$\eqeel$ via:
\begin{equation}
\jzero \eqeel(E) = q \eqepv(E) \phibb(E)
\label{eq:j0}
\end{equation}
Here, $E$ is the photon energy and $\phibb$ the black body spectrum at a given temperature.\cite{Vandewal2009} It follows from Eq.~(\ref{eq:voc}) and~(\ref{eq:j0}) that~$\voc$ is maximized when~$\jzero$ is in its thermodynamic limit, that is, when~$\eqeel$ equals unity and all recombination is radiative.

As compared to the ideal situation in the Shockley--Queisser model, typically three loss channels are considered in OPVs. First, energetic driving force due to an energy level offset between electron donor and acceptor; even though these losses have been drastically reduced by the transition from fullerene to nonfullerene acceptors~(NFAs),\cite{PerdigonToro2020,Liu2016} they may still amount to 0.2--\unit[0.3]{V} depending on the local energy landscape at the donor/acceptor~interface.\cite{Nakano2019} Second, nonradiative recombination, reducing $\voc$ by $kT/q \ln(\eqeel)$, which is estimated to be at least about~\unit[0.2]{V} even in the best OPVs to date.\cite{Benduhn2017,Liu2020} Third, energetic disorder in the density of states~(DOS); it has been shown that the higher the disorder~$\sigma$, the lower the~$\voc$, since carriers sink deeper into the DOS.\cite{Blakesley2011}

The concepts outlined above have in common that they---either explicitly or implicitly---assume photogenerated carriers to be in thermal equilibrium with the lattice. This assumption is well justified for inorganic semiconductors, where band transport dominates and thermalization occurs by phonon emission on sub-ps timescales, such that carriers are transported at quasi-equilibrium energies.\cite{Wurfel2000} In organic materials, relaxation is more complicated and consists of two distinct processes. First, a fast, sub-picosecond thermalization by coupling to molecular vibrations brings the system to the lowest excited state.\cite{Nemec2009,Lane2015} Since the local site energy is typically not the global energy minimum, a second thermalization occurs via thermally activated tunneling~(hopping) in the typically broad distribution of localized sites, which is slow.\cite{Bassler1993} Experimental and numerical studies have shown that excess carriers in OPVs are collected before this second process has completed, i.e., before photogenerated charges are fully relaxed in their respective DOS.\cite{Melianas2015,Melianas2017} Also the distribution of charge transfer~(CT) states that form under steady state illumination was shown to be characterized by an effective temperature that exceeds that of the ambient.\cite{Brigeman2018} Recent work provides evidence that also the EL of the interfacial CT~state is governed by nonequilibrium effects.\cite{Melianas2019} However, surprisingly little is known about how slow thermalization in OPVs affects the device~$\voc$.

Here, we show that the~$\voc$ of disordered OPVs can significantly exceed its equilibrium value. We demonstrate this for two material systems, a traditional polymer:fullerene blend and a recent polymer:NFA blend with 16\%~efficiency. In both cases, the experimental~$\voc$ is 0.1--\unit[0.2]{V} higher than predicted by quasi-equilibrium device simulations and by Eq.~(\ref{eq:voc}) with input parameters from the reciprocity analysis.\cite{Rau2007,Kirchartz2016} Instead, using an experimentally calibrated kinetic Monte Carlo~(KMC) model\cite{Wilken2020} gives a good description of the device~$\voc$, as well as its dependence on thickness and temperature. With this, we show that the excess energy due to incomplete thermalization can actually be harvested and we propose that disordered OPVs can work as `hot' carrier solar cells.

\begin{figure*}[t]
\includegraphics[]{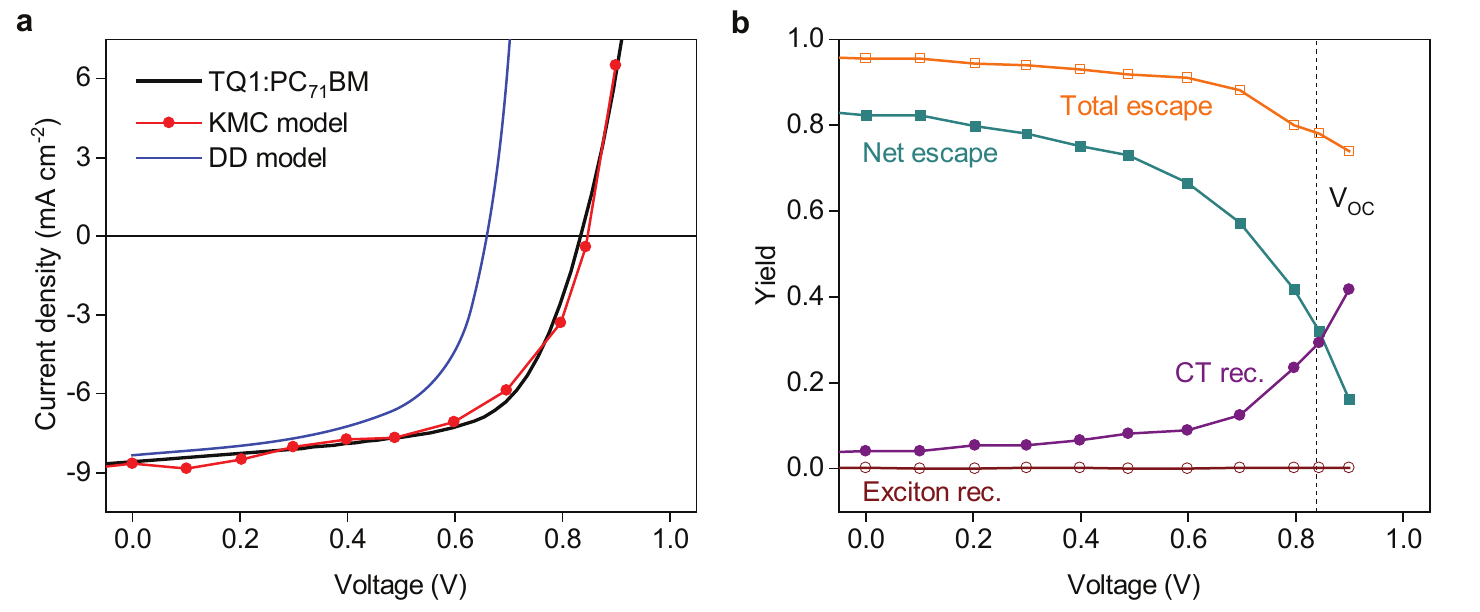}
\caption{\textbf{Measured versus modeled $\boldsymbol{J}$--$\boldsymbol{V}$ curves and loss analysis.} (a)~The black line represents measured $J$--$V$ characteristics of a 70-nm thick TQ1:P\ce{C71}BM solar cell under illumination. Only the KMC model~(red symbols) gives an accurate description of the experiment, while the DD~model~(blue line) that inherently assumes thermal equilibrium underestimates~$\voc$ by about~\unit[0.2]{V}. KMC and DD models use a single, consistent set of parameters. (b)~Corresponding extraction and loss yields from KMC. Total and net escape yields are defined as $y_\text{total} = (J_{n,\text{an}} + J_{n,\text{cat}} + J_{p,\text{an}} + J_{p,\text{cat}})/J_\text{abs}$ and $y_\text{net} = (- J_{n,\text{an}} + J_{n,\text{cat}} + J_{p,\text{an}} - J_{p,\text{cat}})/J_\text{abs}$, where $J_{(n/p),(\text{an}/\text{cat})}$ is the current density of photogenerated electrons/holes extracted via the anode/cathode and $J_\text{abs}$ is the current density corresponding to light absorption. The curves labeled exciton and CT~recombination show the relative current densities associated with exciton and CT~recombination, i.e., the fraction of photogenerated charges that undergo these processes. Similar data for the DD~model can be found in the Supporting Information~(Section~9).}
\label{fig:figure1}
\end{figure*}

We will first focus on TQ1:P\ce{C71}BM blends, for which the importance of nonequilibrium effects is well documented.\cite{Melianas2015,Melianas2017,Melianas2019} Recently, we have developed a KMC~model that can describe current-voltage~($J$--$V$) curves of complete OPVs and predict the device~$\voc$, $\jsc$ and~FF under illumination.\cite{Wilken2020} Figure~\ref{fig:figure1} demonstrates this for a TQ1:P\ce{C71}BM solar cell with an active-layer thickness of~\unit[70]{nm}. The model fully accounts for slow relaxation in the disorder-broadened DOS~(assumed to be Gaussian in shape) and makes experimentally justified assumptions about the carrier injection at the contacts and the CT~recombination rate. Furthermore, it implements a minimalistic but sufficiently realistic model of the morphology, consisting of a molecularly mixed TQ1:P\ce{C71}BM matrix with embedded P\ce{C71}BM aggregates.\cite{Wilken2020} The model and the used parameters are further discussed in the Supporting Information~(Section~2). 

It is, in this context, important that KMC is the gold standard for charge transport simulations in this type of materials. Although drift--diffusion~(DD) models can reproduce KMC~results in certain cases, like space-charge limited transport,\cite{Pasveer2005} when the right mobility functionals and boundary conditions are used, DD is a simplification that, amongst others, upfront assumes charge carrier populations to be fully thermalized. Hence, to find out how the simplifications in DD work out in the case of OPVs, we calibrated the KMC~model to the experiments and used the obtained parameters as input for a DD~model. Specifically, we aimed to describe the $J$--$V$~curve with a drift--diffusion~(DD) model that assumes photogenerated electrons and holes to be in thermal equilibrium. The input parameters of the DD~model were fully consistent with those of the KMC~model. In particular, we assumed identical injection barriers and the same energy gap between the highest occupied molecular orbital~(HOMO) of the donor and the lowest unoccupied molecular orbital~(LUMO) of the acceptor, while energetic disorder was implemented via established mobility functionals.\cite{Pasveer2005} In other words, both models are used to describe the very same device, with the crucial difference that in the KMC~model, nonequilibrium effects are accounted for, whereas the DD~formalism is inherently based on the assumption of near-equilibrium through the use of Boltzmann statistics. In the limit that nonequilibrium effects are unimportant, DD~simulations with parameterized mobilities, as used here, accurately reproduce the more detailed KMC~calculations.\cite{Pasveer2005} Full details of the simulations are provided in the Supporting Information~(Section~2). As can be seen from the blue line in Figure~\ref{fig:figure1}, the DD~model describes $\jsc$ and the shape of the $J$--$V$ curve reasonably well, but significantly underestimates $\voc$ by about~\unit[0.2]{V}. This clearly shows that the equilibrium DD~approach does not capture all relevant physics.

\begin{figure*}[t]
\includegraphics[width=\linewidth]{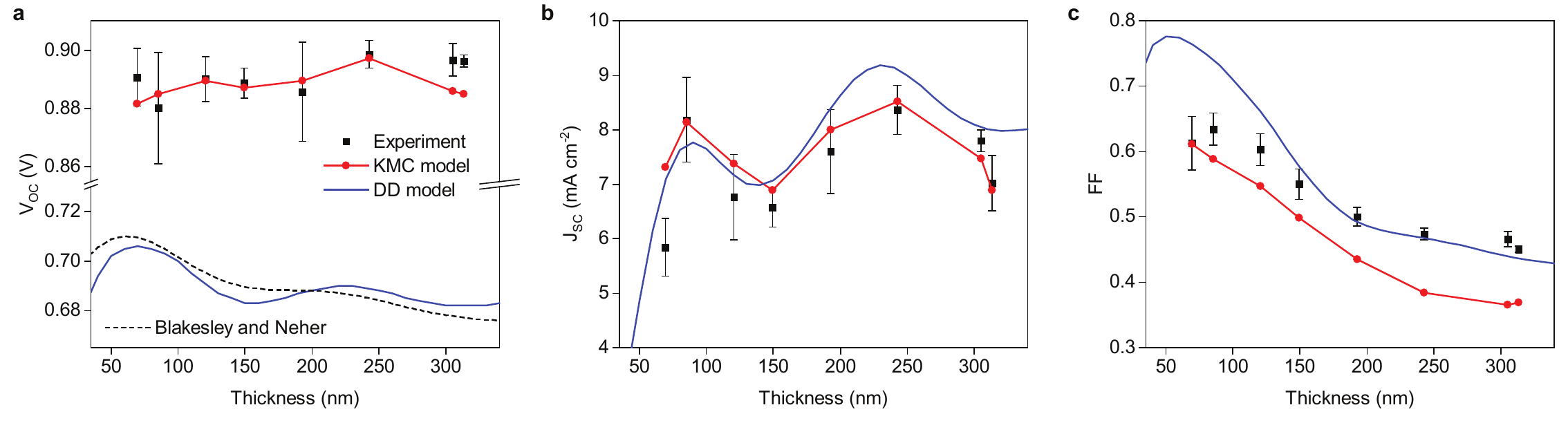}
\caption{\textbf{Thickness-dependent device performance.} Shown is the (a)~open-circuit voltage, (b)~short-circuit current and (c)~fill factor of TQ1:P\ce{C71}BM solar cells with varying active-layer thickness~(black symbols) together with the predictions of the KMC~model~(red symbols) and the DD~model~(blue lines). The dashed line in panel a represents the analytical model by Blakesley and Neher\cite{Blakesley2011} assuming carriers to be fully relaxed.}
\label{fig:figure2}
\end{figure*}

To test the general validity of our statement and to highlight the predictive value of our KMC~model, we fabricated OPVs with varying active-layer thickness. Thickness variations allow to probe the device characteristics for a range of extraction times and carrier densities, and intuitively, one might expect nontrivial behavior for systems where extraction competes with thermalization. The variation of the generation rate profile with thickness was explicitly taken into account via transfer-matrix modeling.\cite{Burkhard2010} Figure~\ref{fig:figure2} shows that while the KMC~model provides an excellent description of the device~$\voc$, the mismatch between experiment and DD~model remains independent of the thickness. The trend in~$\voc$ predicted by the DD~model is in good agreement with the analytical model by Blakesley and Neher,\cite{Blakesley2011} which supposes full relaxation to effectively reduce the HOMO--LUMO gap by $\sigma^2/kT$. Specifically, in the DD model~$\voc$ follows the modulation in charge generation rate~(and thus, $\jsc$) due to interference in the multilayer device, as suggested by Eq.~(\ref{eq:voc}). This strict correlation is neither observed in the experiment, nor in the KMC~calculation, which suggests that~$\voc$ actually results from a more complex interplay between generation~(profiles) and thermalization. The absence of a clear decreasing trend in~$\voc$ with increasing thickness might appear at odds with the notion of an ongoing and incomplete thermalization. However, thermalization in disordered media follows roughly a log--linear time dependence till an equilibrium energy is reached.\cite{Bassler1993} The implication of that is that the differences in extraction time for experimentally achievable thickness variations only cause minor differences in thermalization and are overwhelmed by other effects like changes in absorption and recombination.

In contrast to~$\voc$, both models show a good agreement with the experimental~$\jsc$ and, to a lesser degree, the FF, see Figure~\ref{fig:figure2}b and c, respectively. Hence, the kinetic competition between extraction and recombination is also reasonably described by the DD~model.\cite{Bartesaghi2015,Neher2016} This is especially the case for the thick devices, where the FF approaches its space-charge limit due to imbalanced transport.\cite{Wilken2020b,Mihailetchi2005} We note that the limitations of DD~modeling regarding $\voc$ can be partly overcome without loosing accuracy in~$\jsc$ and FF by artificially increasing the band gap, but this would be an ad hoc compensation of the inability of the DD~model to capture nonequilibrium effects, and lead to an inconsistency between the parameters in the two models. The fact that the band gap effectively acts as a fit parameter provides a plausible explanation why DD~modeling is so successful in describing $J$--$V$~curves of OPVs. It also confirms that it can indeed lead to useful and valid results when looking at certain macroscopic phenomena such as space charge or carrier injection and extraction. However, the point we want to make here is that when dealing with questions about the nature and limiting factors of~$\voc$, DD~models provide an incomplete picture of the physical reality in OPVs.

We recently demonstrated that our KMC~model can also be used to accurately describe the spectral shape and position of the solar cell's absorption and emission.\cite{Melianas2019,Felekidis2020} This gives us the opportunity to estimate~$\voc$ from Eq.~(\ref{eq:voc}) with the KMC~input parameters used in Figure~\ref{fig:figure1}, assuming strict equilibrium conditions. The procedure is detailed in the Supporting Information~(Section~3) and gives~$\voc \approx \unit[0.69]{V}$. Despite the simplifications made, this value is strikingly close to the value of~$\voc \approx \unit[0.66]{V}$ found in the DD~simulations implicitly assuming Boltzmann statistics, that is near-equilibrium conditions, are implicitly assumed. Hence, the DD and the analytical model consistently show that~$\voc$ would be $\sim$\unit[0.2]{V} lower than actually measured if the device was operating in thermal equilibrium. In other words, voltage losses in this OPV~system would be significantly larger if thermalization would complete in the charge carrier lifetime---the device operates as a hot carrier solar cell.\cite{Ross1982,Wurfel1997,Kahmann2019} In the Supporting Information~(Section~4), we confirm the result from our earlier work\cite{Melianas2015} that the charge carrier populations do not reach equilibrium prior to extraction.

Another indication that far-from-equilibrium charges contribute significantly to the $J$--$V$~curve under illumination comes from the incomplete saturation of the photocurrent at short circuit~($V = 0$). This is just visible in Figure~\ref{fig:figure1}a and more clearly in Figure~S7 in the Supporting Information, where we also show that the KMC~model does and the DD~model does not reproduce the observation. The cause for this difference becomes clear from the loss analysis in Figure~\ref{fig:figure1}b and the corresponding data in Figure~S7. First, the nearly complete saturation of the CT recombination yield already at small forward bias rules out field-dependent charge generation as an explanation.\cite{Mihailetchi2004} Instead, the large difference between net and total escape yields at short-circuit conditions indicates that, despite the $\sim$\unit[1]{V} built-in field, a large fraction of photogenerated charges leaves the device via the wrong (nonselective) contact at short circuit. This points to highly diffusive, nonequilibrium charge motion, requiring large fields to be suppressed. The strong voltage dependency of the difference between net and total escape yields confirms this notion. Such a diffusion-loss scenario is fully in line with our earlier analysis of ultrafast transport and absorption.\cite{Melianas2015,Melianas2017,Melianas2019}

\begin{figure*}[t]
\includegraphics[width=0.6\textwidth]{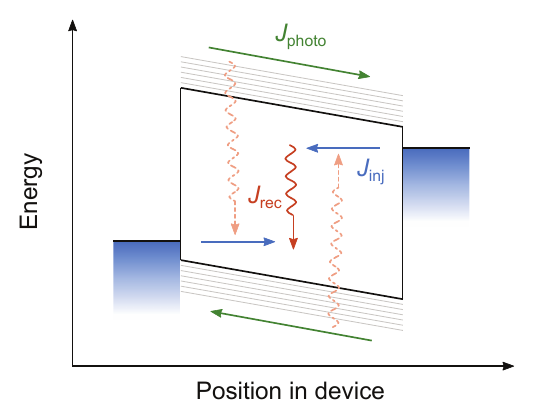}
\caption{\textbf{Schematic charge carrier kinetics in an operational OPV~device at open-circuit conditions.} Nonthermalized photocurrents~($J_\text{photo}$, green arrows) and thermalized injection currents~($J_\text{inj}$, blue arrows) balance. Recombination~(corresponding current: $J_\text{rec}$) is weak for photogenerated charges but strong for injected charges. The gradual but incomplete thermalisation of the photocharges is not shown.}
\label{fig:figure3}
\end{figure*}

Another important conclusion that can be drawn from the total and net escape yields in Figure~\ref{fig:figure1}b is that at~$\voc$, about~80\% of the photogenerated charges do not recombine, but instead are extracted from one of the contacts, c.f.~total escape yield. Since $\sim$30\% of the photogenerated charges are extracted at the desired contact, c.f.~net escape yield, there is a net photocurrent. Since $J = 0$ at~$\voc$, there must be a balancing injection current corresponding to $\sim$30\% of the short-circuit current. The existence of such an injection current is hard, if not impossible to measure directly. However, a strong indication for this happening comes from the recent observation that recombination kinetics around~$\voc$ in this system follow near-equilibrium rates:\cite{Roland2019} since under $\voc$~conditions the vast majority of `nonequilibrium' photogenerated charges still leaves the device via the contacts, recombination is likely to be governed by `equilibrium' charges being injected from the contacts that act as thermal reservoirs. On top of this comes the small fraction of `thermalized' photogenerated charges that does get trapped in the disordered~DOS. The above considerations are condensed in the cartoon shown in Figure~\ref{fig:figure3} and are further discussed in Supporting Information~(Section~4) where it is shown that even at open-circuit conditions, a high-energy (nonthermalized) photocurrent runs through the device, as argued above and indicated in Figure~\ref{fig:figure3}.

\begin{figure*}[t]
\includegraphics{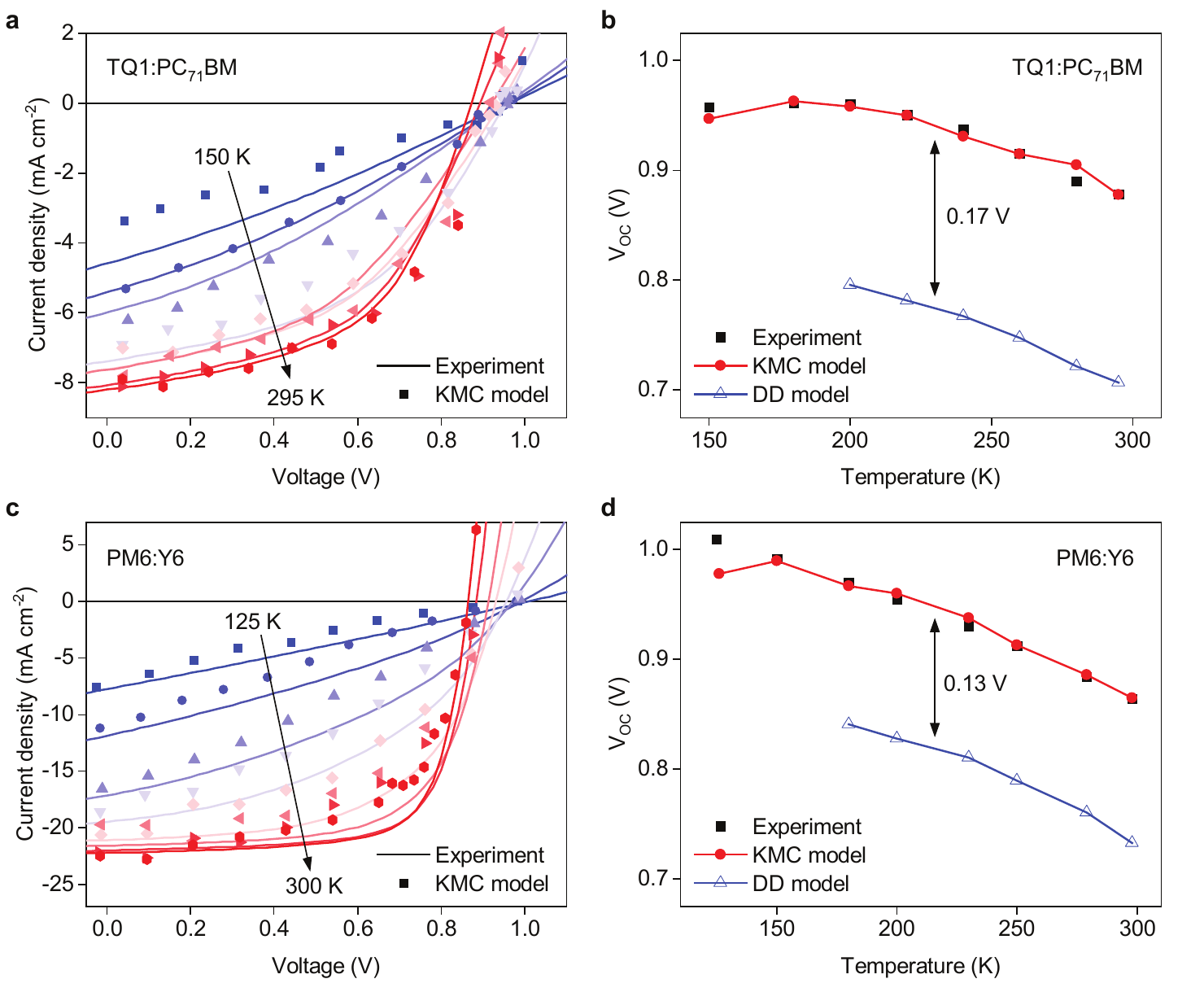}
\caption{\textbf{Temperature-dependent device performance.} Panels~(a) and~(c) show temperature-dependent $J$--$V$ curves for a \unit[75]{nm} thick TQ1:P\ce{C71}BM solar cell and a \unit[115]{nm} thick PM6:Y6 solar cell, respectively. Solid lines are experimental data and symbols KMC~simulations. Panels~(b) and~(d) show the corresponding values of~$\voc$ and compare them with the results of the DD~model.}
\label{fig:figure4}
\end{figure*}

The excess energy of photogenerated charges due to the incomplete thermalization during extraction forms a reservoir of energy. Equivalently, it can be considered as a (time-dependent) effective temperature of the charge carrier distribution that significantly exceeds the lattice temperature..\cite{Marianer1992} This introduces another relevant energy scale in the system, and one may anticipate a pronounced effect on the temperature dependence of the device characteristics. Figure~\ref{fig:figure4}a,b shows the temperature-dependent $J$--$V$~behavior for a relatively thin TQ1:P\ce{C71}BM~device. Clearly, the KMC~model~(using the same parameters as before) provides a reasonable description of the experiment at all temperatures, confirming that it captures all essential device physics. This is in stark contrast to the DD~model, which not only underestimates~$\voc$ regardless of temperature, but also  fails to capture the temperature dependence of the shape of the $J$--$V$~curves, as shown in the Supporting Information, Figure~S8. Nevertheless, the slope of~$\voc$ versus temperature is very comparable in both KMC and DD~simulations, and the same holds for the forward part of the $J$--$V$~curves. We attribute this to fact that in both models the temperature dependence of~$\voc$ is dominated by the strong temperature dependence of the forward injection current that, as discussed in the context of Figure~\ref{fig:figure3}, is carried by a charge carrier population that reflects the equilibrium temperature.

So far, we have shown that a relatively amorphous and disordered polymer:fullerene OPV~system should be treated as a far-from-equilibrium hot carrier device. An important question is whether or not the same holds true for modern NFA~systems. To this end, we performed a similar series of experiments on the state-of-the-art PM6:Y6~system that provides power conversion efficiencies over 15\%.\cite{Yuan2019,PerdigonToro2020} In a recent work,\cite{Upreti2019} we investigated the energetic disorder in a range of OPV~blends, and found Gaussian disorder values~$\sigma$ ranging from~45 to~\unit[80]{meV}, without any clear correlation between~$\sigma$ and~$\voc$. For the PM6:Y6~system, $\sigma_\text{HOMO} \approx \unit[89]{meV}$ and $\sigma_\text{LUMO} \approx \unit[68]{meV}$ were found. As these numbers are much higher than the thermal energy~$kT \approx \unit[25]{meV}$ at room temperature and well exceed the thresholds we previously found for disorder to become negligible,\cite{Melianas2017} one should expect that the phenomena discussed above are relevant for this NFA~system as well. Figure~\ref{fig:figure4}c,d shows the temperature dependence for a \unit[115]{nm} thick PM6:Y6~solar cell. Although the difference with the KMC~model is smaller than for the more disordered TQ1:P\ce{C71}BM~system, also the highly efficient PM6:Y6~system cannot be correctly described by the DD~model, see also Figure~S9 in the Supporting Information, while the KMC~model reproduces the experiment well, especially regarding the temperature dependence of~$\voc$. Again, consistent input parameters were used for DD and KMC~modeling, see the Supporting Information~(Section~2). This shows that also for NFA~systems with relatively low disorder and balanced electron and hole mobilities, $\voc$ is insufficiently described by equilibrium concepts. A further discussion of the role of disorder on the $\voc$~difference between KMC and DD is given in the Supporting Information~(Section~8). 

\begin{figure}[t]
\includegraphics{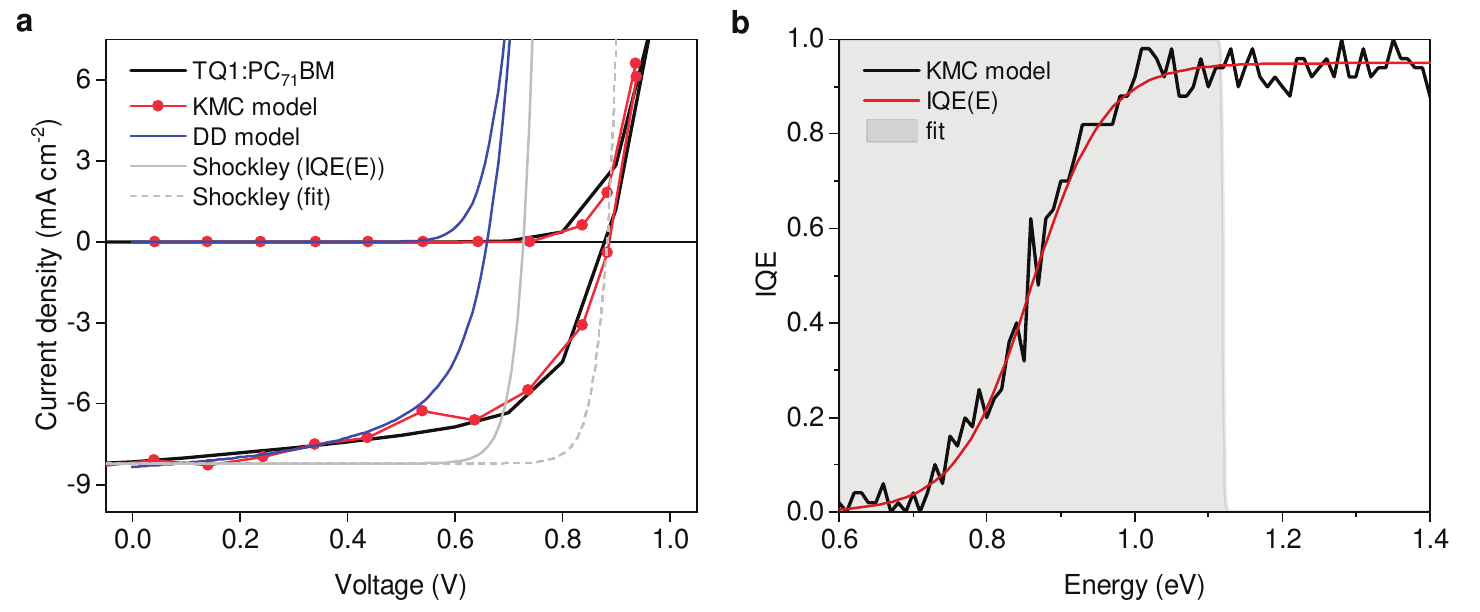}
\caption{\textbf{Analysis of difference between DD and KMC models.} (a)~$J$--$V$~curves in dark and light~(solid black lines) compared to DD~(blue lines) and KMC~(red symbols). The shown sample is nominally identical to that in Figure~\ref{fig:figure1}, the simulations are the same but slightly rescaled to match~$\jsc$. The grey lines are calculated from the Shockley equation, $J = J_0 (\exp(qV/kT) - 1)$, with different~$J_0$ as explained in the text. (b)~IQE calculated by KMC~(black line) and empirical~$\text{IQE}(E)$ function~(red line). The grey shaded area is explained in the text.}
\label{fig:figure5}
\end{figure}

As shown in Figure~\ref{fig:figure5}, a remarkable difference between KMC and DD~simulations is in the dark currents that show an upswing that is shifted by essentially the same amount as~$\voc$, despite the fact that the very same boundary conditions are used, see the Supporting Information~(Section~2). This difference cannot be attributed to nonequilibrium effects since in the dark all charges are injected from thermalized reservoirs instead of being photogenerated. The dark $J$--$V$~curves are consistent with those under illumination in the sense that the superposition principle~$J_\text{light} \approx J_\text{dark} - \jsc$ is obeyed. Approximating the dark current with the Shockley equation, $J_\text{dark} = J_0 (\exp(qV/kT) - 1)$, this means that the dark reverse saturation current~$J_0$ must be different in both models. Above we discussed how for DD~$J_0$ and concomitantly~$\voc$ can be calculated from the model input parameters as the overlap of the $\eqepv$~spectrum with the black body spectrum, see also the Supporting Information~(Section~3). To save a near-equilibrium interpretation of~$\voc$ in case of KMC, one would have to find a meaningful alternative way of calculating~$J_0$. In the used framework, the most logical and in fact only viable way is to assume that part of the absorption spectrum does not contribute to the EQE, which implies an IQE that drops to zero below some threshold energy. For the TQ1:P\ce{C71}BM~system, such behavior has indeed been observed.\cite{Felekidis2020} However, to reproduce the~$\voc = \unit[0.88]{V}$ value from KMC, one would have to cut the absorption spectrum below~$\sim$\unit[1.12]{eV}. Doing so produces the dashed grey line in Figure~\ref{fig:figure5}a. However, following the method laid out in our earlier work,\cite{Felekidis2020} we calculated the energy-dependent IQE for the parameters used, see Figure~\ref{fig:figure5}b. Clearly, the true IQE spectrum does not roll off at~\unit[1.12]{eV} but around~\unit[0.86]{eV}. Using the empirical function shown by the red line as~$\text{IQE}(E)$ leads to a~$\voc$ around~\unit[0.73]{V}, as shown by the solid grey line in Figure~\ref{fig:figure5}a, which is somewhat higher than found for unity IQE, but still far of the actual~$\voc$. The observed inconsistency between the fitted IQE spectrum and the actual IQE spectrum~(grey area in Figure~\ref{fig:figure5}b) points towards a fundamental problem in the used reciprocity formalism when applied to OPVs. Specifically, the assumed equivalence of injection and extraction will be violated when one of the channels is far from equilibrium while the other is not.\cite{Kirchartz2016}

The above strong indications for the importance of nonequilibrium kinetics for the performance of two prototypical OPV~systems naturally raises the question why previous analysis in terms of reciprocity relations, which are derived on the assumption that detailed balance holds, worked so well.\cite{Rau2007,Vandewal2009,Roland2019,Benduhn2017,Vandewal2014,Qian2018} Here, one has to make a distinction between at least two different uses of reciprocity relations. First, in many works reciprocity between emission and action spectra is used to convert one into the other with the aim to establish an estimate for the `relaxed' CT energy, which subsequently is used as a reference point for further analysis.\cite{Vandewal2014,Benduhn2017,Vandewal2009,Qian2018} In our previous work, we have shown that this reciprocity is not strictly obeyed due to the CT~emission coming from a nonthermal subset of the full CT~manifold.\cite{Melianas2019,Felekidis2020} However, the typical energy range over which experiments can be analyzed makes it hard to pinpoint systematic deviations from the phenomenological reciprocity.

More interesting in the current context is the use of Eq.~(\ref{eq:voc}) and~(\ref{eq:j0}) to predict~$\voc$. It was shown by Roland~et~al.,\cite{Roland2019} for example, that experimentally measured~$\eqeel$ and~$\eqepv$ spectra can be used to accurately predict the open circuit voltage of 150--\unit[200]{nm} thick TQ1:P\ce{C71}BM~devices. Although this topic warrants further investigation, we speculate that at least part of the answers is due to a cancelation of errors. In Ref.~\citenum{Roland2019}, $\voc$ is calculated as $\voc = \vocrad + (kT/q) \ln(\eqeel)$, using an~$\eqeel$ in the range of~$10^{-5}$--$10^{-6}$, which is a common value for OPV~systems and corresponds to a voltage loss around 0.3--\unit[0.35]{V}.\cite{Qian2018} As also argued in the Supporting Information~(Section~3), the reasonable assumption that CT and S$_1$ recombination are competing against the same or at least similar loss channels, yields that their relative lifetimes, and therefore their relative~$\eqeel$, should reflect their relative oscillator strengths. Interestingly, for the few OPV~systems for which CT~lifetime estimates are known, it differs by 1--2~orders of magnitude from the S$_1$~lifetime that is typically in the ns~range in OPVs.\cite{Mikhnenko2015} This ratio is consistent with the typical difference in CT and S$_1$ absorption strengths,\cite{Vandewal2014,Roland2019,Felekidis2020} but not with an~$\eqeel$ for CT~recombination of~$10^{-5}$--$10^{-6}$. Using instead an~$\eqeel$ around~$10^{-2}$, i.e., the approximate ratio of the~S$_1$ and CT~lifetime, leads to a voltage loss of $\sim$\unit[0.12]{V}, which happens to differ from the original 0.3--\unit[0.35]{V} voltage loss by an amount that is rather similar to the $\sim$\unit[0.2]{V} difference in~$\voc$ that we found between our KMC~simulations and the reciprocity prediction in the Supporting Information~(Section~3). A concise overview of the differences between equilibrium and nonequilibrium interpretations of~$\voc$ is given in the same section of the Supporting Information.

A possible partial explanation for the severe underestimation of the~$\eqeel$ in experiments is that it is tacitly assumed that all injected charge at voltages corresponding to~$\voc$ recombines. This need not be the case in carefully optimized OPV~morphologies that consist of phase separated donor and acceptor domains. In such morphologies, which are designed to efficiently keep electrons and holes apart, injected electron and hole currents might very well never meet. This is especially the case in blends where mixed and pure regions coexist, resulting in an energy cascade that pushes charge carriers away from the donor/acceptor interface.\cite{Burke2014} In our KMC~simulations for the TQ1:P\ce{C71}BM~system, this recombination fraction is~$\sim$$10^{-1}$, which is an upper limit since the used simplified morphology leads to a significant underestimation of the fill factor for thicker devices~(Figure~\ref{fig:figure2}c).

Summarizing, we have shown for two exemplary OPV systems, a polymer:fullerene and polymer:NFA blend, that $\voc$ significantly exceeds its equilibrium value by 0.1--\unit[0.2]{V}. The excess energy arises because charge carriers are not completely relaxed in their disorder-broadened~DOS when they are extracted at the contacts. Our results indicate that even under $\voc$~conditions, most of the photogenerated charge carriers do not recombine, but leave the device via one of the contacts. Instead, recombination is largely dominated by thermalized injected carriers, which explains the success of equilibrium concepts in the past. It should be noted that the higher~$\voc$ does not necessarily translate into higher efficiency, as the latter depends on all three of the parameters~$\voc$, $\jsc$ and FF. Whether the nonequilibrium effects can be exploited to break the Shockley--Queisser limit or to realize OPVs with significantly higher film thicknesses is an interesting direction for future research.\cite{Andersson2020} Since the material systems investigated here are not exceptional in terms of energetic disorder, but typical representatives of the state of the art, we expect that our results are highly relevant for most OPV systems.

\section{Acknowledgements}
We thank Dr.~Ergang Wang at Chalmers University of Technology, G\"oteborg~(Sweden), for synthesizing the~TQ1, Dr.~Jun Yuan at Central South University, Changsha~(China), for synthesizing the Y6 and Prof.~Feng Gao for stimulating discussions. T.U.~acknowledges financial support by the Swedish Research Council~(project `OPV2.0'). This project has received funding from the European Union's Horizon 2020 research and innovation programme under the Marie Sk\l{}odowska-Curie grant agreement No~799801~(`ReMorphOPV'). M.K.~thanks the Carl Zeiss Foundation for financial support. 

\section{Author contributions}
T.U.~performed and analyzed all experiments and fabricated the TQ1:P\ce{C71}BM devices. T.U.~and S.W.~performed the KMC and DD~simulations. H.Z.~fabricated and pre-characterized the PM6:Y6~devices. M.K.~wrote the simulation software, conceived the idea and coordinated research. S.W.~and M.K.~wrote the manuscript with input from T.U. All authors contributed to discussions.

\bibliography{ms}

\end{document}


\clearpage

\tableofcontents

\clearpage

\pagebreak
\section{Experimental Details}

\begin{figure}
\vspace{1mm}
\includegraphics[width=0.8\textwidth]{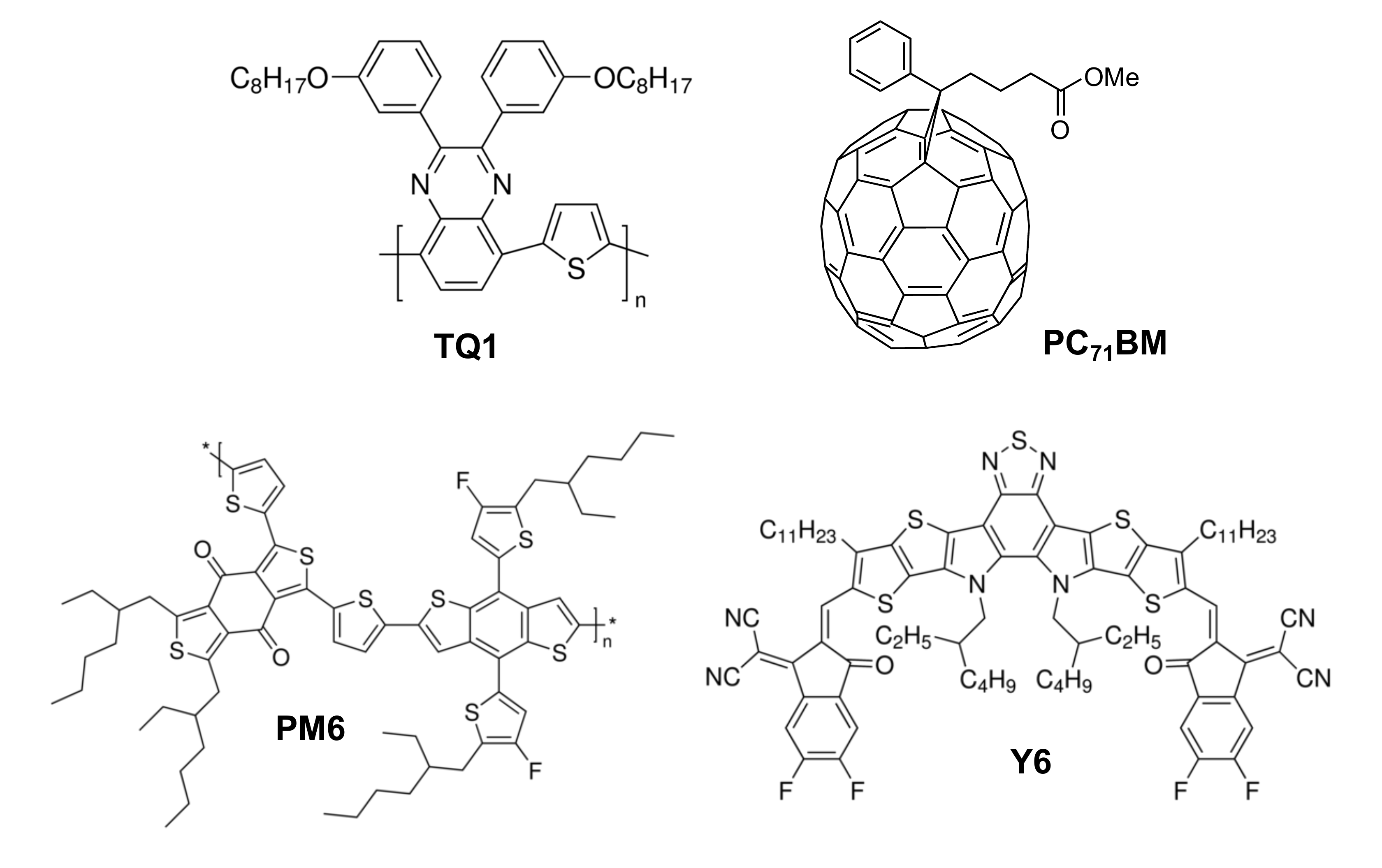}
\caption{Chemical structures of the photoactive materials used in this study.}
\label{fig:figS1}
\end{figure}

\paragraph{Materials} The chemical structures of the photoactive materials used in this study are shown in Figure~\ref{fig:figS1}. The poly[[2,3-bis(3-octyl\-oxy\-phenyl)-5,8-quinoxa\-line\-diyl]-2,5-thio\-phene\-diyl]~(TQ1) polymer was synthesized as described previously.\cite{Wang2010} [6,6]-phenyl-\ce{C71}-butyric acid methyl ester~(\ce{PC71BM}) was purchased from 1-Material. Poly[(2,6‐(4,8‐bis(5‐(2‐ethyl\-hexyl‐3‐fluoro) thio\-phen‐2‐yl)‐benzo[1,2‐b:4,5‐b$'$]di\-thio\-phene))‐alt‐(5,5‐(1$'$,3$'$‐di‐2‐thienyl‐5$'$,7$'$‐bis (2‐ethyl\-hexyl)benzo[1$'$,2$'$‐c:4$'$,5$'$c$'$]di\-thio\-phene‐4,8‐dione~(PM6) was purchased from Solarmer Materials. 2,2$'$‐((2Z,2$'$Z)‐((12,13‐bis(2‐ethyl\-hexyl)‐3,9‐di\-undecyl‐12,13‐di\-hydro‐[1,2,5]thi\-adi\-azolo[3,4‐e]thieno[2,$''$30$''$:4$'$,5$'$]thieno[2$'$,3$'$:4,5]pyrrolo[3,2‐g]thieno[2$'$,3$'$:4,5]thieno[3,2‐b]indole‐2,10‐diyl)bis(methanyl\-ylidene))bis(5,6‐di\-fluoro‐3‐oxo‐2,3‐di\-hydro‐1H‐indene‐2,1‐di\-yl\-ide\-ne))di\-ma\-lono\-nitrile(Y6) was synthesized according to literature.\cite{Yuan2019}

\paragraph{Device fabrication} TQ1:P\ce{C71}BM solar cells were fabricated with the structure indium tin oxide (ITO)/poly(3,4-ethylene\-dioxy\-thio\-phene) poly\-styrene sulfonate (PEDOT:PSS)/active layer/LiF/Al. The PEDOT:PSS aqueous solution~(Baytron P VP AI 4083) was spin-coated on cleaned ITO-covered glass substrates at \unit[3000]{rpm} for \unit[15]{s}, followed by annealing at \unit[150]{$^\circ$C} for \unit[10]{min} in air, to form a \unit[30]{nm} film. The active layer was spin-coated at \unit[500]{rpm} for \unit[60]{s} from a 1:2.5~(w/w) blend solution of TQ1 and P\ce{C71}BM in chlorobenzene. The concentration of the solution was varied from 20 to \unit[100]{mg\,ml$^{-1}$} to yield an active-layer thickness of 70--\unit[310]{nm}. Finally, the top electrode of LiF~(\unit[0.6]{nm}) and Al~(\unit[90]{nm}) was thermally evaporated under high vacuum. PM6:Y6 solar cells were fabricated with the structure ITO/PEDOT:PSS/active layer/PFN-Br/Al. The active layer was spin-coated at 2500--\unit[6000]{rpm} from a 1:1.2~(w/w) blend solution of PM6 and Y6 in chloroform with 0.5\% 1-chloronaphthalene as an additive. Subsequently, the electron-extracting layer of PFN-Br~(Solarmer Materials) was spin-coated from a \unit[0.5]{mg\,ml$^{-1}$} solution in methanol. Finally, the top electrode was evaporated under high vacuum. 

\paragraph{Characterization} Film thicknesses were determined using a Dektak~6M stylus profilometer. The thicknesses of the devices were averaged over 4~samples each with 4~pixels. Room temperature current--voltage curves were recorded with a Keithley 2400 source measure unit under standard AM1.5G illumination~(\unit[100]{mW\,cm$^{-2}$}) using an Oriel LSH-7320 solar simulator. For temperature-dependent measurements, TQ1:P\ce{C71}BM devices were mounted in a liquid‐nitrogen cryostat controlled by a Lake Shore 330 temperature controller. The assembly had a built-in solar simulator with a Xe-arc lamp which was calibrated for the spectral mismatch. PM6:Y6 devices were tested in a closed cycle helium cryostat~(Advanced Research Systems) and illuminated with a \unit[532]{nm} laser~(Thorlabs). A Keithley 2400 Source Meter provided the voltage and measured current for both the systems. Temperature-dependent measurements were averaged over cooling and heating sweeps to account for minor time dependencies. The monochromatic excitation used for the PM6:Y6 devices is unproblematic for the current study that focusses on hot-carrier~(distribution) effects. Since hot-carrier phenomena are based on the presence of excess photon energy, the condition for these to occur is that the difference between the photon energy and the energy gap of the absorber is significant. For the \unit[532]{nm}~(\unit[2.33]{eV}) laser, the condition is easily met for both PM6 and Y6 that have an optical energy gap around 600 and \unit[900]{nm}, respectively.\cite{Yuan2019} In addition the `hotness' of the charge carrier populations stems to a large degree from the energy gained upon charge transfer at the donor/acceptor interface, which happens as long as one does not selectively excite deep in the CT state.\cite{Felekidis2020}

\pagebreak
\section{Details of the Numerical Models}

\subsection{Kinetic Monte Carlo Model}
The kinetic Monte Carlo~(KMC) model is implemented on a simple cubic grid such that the nearest neighbor hopping distance $a_\text{NN}$ equals the lattice constant and relates to the total site density~$N_0$ as~$a_\text{NN} = N_0^{-1/3}$. While the code allows to consider hopping to different numbers of neighbor sites, we used strict nearest neighbor hopping here to warrant consistency with earlier work. Below the model and its parameters are explained in detail.

\paragraph{Hopping Rates} We use the Miller-Abrahams expression to quantify, with the least number of parameters, the nearest-neighbor hopping rate of a charge carrier from an initial state~$i$ with energy~$E_i$ to a final state~$f$ with energy~$E_f$ as
\begin{equation}
\nu_{ij} = \nu_0\exp(-\alpha r_{if}) \begin{cases}
    \exp\left(-\frac{E_f - E_i \pm q \vec{r}_{if} \cdot \vec{F} + \Delta E_C}{kT}\right)& \Delta E > 0\\
    1              & \Delta E \leq 0
\end{cases}
\label{eq:hopping}
\end{equation}
Here, $\vec{F}$ is the external electric field, $\vec{r}_{if}$ the vector connecting initial and final sites, $\nu_0$  the attempt-to-hop frequency, and $q$ the positive elementary charge. The $+$~($-$) sign refers to electron (hole) hopping. In the configuration used (strict nearest neighbor hopping), the localization length~$\alpha$ is unimportant and the first exponential term of Equation~(\ref{eq:hopping}) was implicitly included in~$\nu_0$, that is, the rate of downward nearest-neighbor hops. In the context of this work, it is important that including non-nearest neighbors as final sites has a similar effect on the relaxation rate as the corresponding mobility increase by an increase in~$\nu_0$. That is, increasing the number of neighbors while keeping the mobility constant does not significantly affect the thermalization process.

\paragraph{Energetics} The term $\Delta E_C$ is the change in Coulomb energy and is calculated by explicit evaluation of the interaction of the moving charge with (a)~all other charges in the simulated device and (b)~their image charges, as well as of the interaction of the image charges of the moving particle with (c)~the particle itself and (d)~all other particles. Image charges arise when metallic contacts are present; the number of image charges accounted for in the simulations is increased till the resulting effective Coulomb potential no longer changes. In order to avoid divergences at zero separation, the Coulomb interaction between a pair of~(unlike) charges, $E_C = -q/(4 \pi \varepsilon_0\varepsilon_r r_\text{eh})$ with $\varepsilon_0\varepsilon_r$ the dielectric constant~($\varepsilon_r = 3.6$) and $r_\text{eh}$ the electron-hole distance, is truncated at minus the approximate exciton binding energy of~$E_b^\text{ex} = \unit[0.5]{eV}$. The single-particle site energies~$E_i$ are drawn from a Gaussian distribution function
\begin{equation}
g(E) = \frac{1}{\sqrt{2\pi\sigma^2}}\exp\left[-\frac{(E - E_0)^2}{2\sigma^2}\right]
\end{equation}
with $E_0$ the mean energy and~$\sigma$ the broadening of the total density of states~(DOS)~$N_0$. The HOMO and LUMO energy of a single site are assumed to be uncorrelated.

\paragraph{Morphology} In previous works, we used an effective hopping medium with~(different) electron and hole hopping parameters~$\nu_0$ and $\sigma$ that correspond to the donor HOMO and acceptor LUMO levels, respectively. The driving force for charge transfer is then implemented via an on-site electron-hole repulsion with a magnitude that equals the LUMO level offset~$\Delta E_\text{LUMO} = \Delta E_\text{LUMO}^D - \Delta E_\text{LUMO}^A$ between donor and acceptor. Here, and in Ref.~\citenum{Wilken2020}, we implemented a simplified phase separated morphology for the TQ1:\ce{PC71BM} system as columnar inclusions~($\unit[7 \times 7]{sites^2}$) in an columnar unit cell~($\unit[10 \times 10]{sites^2}$) where the column axis runs in the current direction~(z). Inclusions were assumed to consist of pure~\ce{PC71BM} with a~\unit[0.2]{eV} lower-lying LUMO compared to the mixed phase; all other properties were left unchanged to keep the number of unknown parameters at a minimum. We did not consider pure TQ1 domains, as our previous electron microscopy experiments do not provide any evidence for them.\cite{Wilken2020} By lack of specific morphological information, the same morphology was used for the PM6:Y6 system.

\paragraph{Excitons} Spatially direct excitons, formed by an electron and a hole on the same site, can recombine with rate~$\nu_\text{ex}$. Similarly, when sitting on neighboring sites, they form a CT~complex that can recombine with rate~$\nu_\text{CT}$. This implies that mono- and bimolecular recombination are treated on equal footing, as recombination rates of exciton and CT~species do not depend on the history of the constituent charges. Exciton diffusion by the Förster resonant energy transfer~(FRET) mechanism is explicitly accounted for. The transition rate is evaluated as
\begin{equation}
\nu_{if}^F = \nu_\text{ex} \left(\frac{R_0}{r_{if}}\right)^6 \Theta\left(E_i^\text{ex} - E_f^\text{ex}\right)
\end{equation}
where $R_0$ is the Förster radius, $\nu_\text{ex}$ the radiative exciton decay rate, $\Theta$ the Heaviside step function, and $E_i^\text{ex}$, $E_f^\text{ex}$ the exciton energies $E_{i/f}^\text{ex} = E_{i/f}^\text{LUMO} - E_{i/f}^\text{HOMO} - E_b^\text{ex}$ at the initial and final sites. Dexter-type exciton diffusion is implicitly accounted for as a double charge hopping process.

\paragraph{Kinetics} The waiting time before an event (hop or recombination) occurs is calculated as
\begin{equation}
\tau = - \frac{\ln(r)}{\Sigma_\nu}
\end{equation}
where $r$ is a random number drawn from a homogeneous distribution between 0 and 1 and~$\Sigma_\nu$ is the sum of the rates of all possible events. The event that occurs after~$\tau$ is selected randomly, using the rates of all possible events as weight factors. Energies, rates, and waiting time are recalculated after each event.

\paragraph{Contacts and Boundary Conditions} Periodic boundary conditions in the x,y-directions were applied for both charge motion and~(image and direct) Coulomb interactions; contacts laying in the z-plane are included unless stated otherwise and were implemented as hopping contacts. We mitigated the `small barrier' problem~(carriers oscillating across the contact interface at large computational cost) by only allowing for a transfer if the number of charges next to the contact interface deviates from its equilibrium value, which is calculated as a Fermi-integral over the actual DOS in the first organic layer next to the contact:\cite{vanderHolst2009}
\begin{equation}
n_\text{cont} = \int_{-\infty}^{\infty} \frac{g(E)}{1 + \exp(E/kT)} \diffd E
\label{eq:Fermi}
\end{equation}
Injection and extraction are modeled as hopping events with an attempt frequency~$\nu_{0,\text{cont}}$ of the same order as for the transport of the faster carrier in the semiconductor. We explicitly checked that this procedure does not limit charge collection or extraction at the contacts for the given parameters and voltages. We also checked that~$\voc$ is not significantly affected by the use of these `buffered' hopping contacts by running a single~$J$--$V$ point with non-buffered hopping at~$\voc$. Both the cathode and anode were considered nonselective; hence, possible losses due to diffusion of carriers into the `wrong' contact are implicitly accounted for.

\begin{table}[t]
\caption{Overview of the parameters used for the kinetic Monte Carlo simulation of TQ1:\ce{PC71BM} and PM6:Y6 solar cells. HOMO and LUMO refer to the orbital energies of the effective medium.}
\begin{tabular}{lccc}
\toprule
& TQ1:\ce{PC71BM} & \multicolumn{2}{c}{PM6:Y6}\\
 \cmidrule(lr){3-4}
Parameter [unit] & Value\cite{Wilken2020} & Value & Literature value\\
\midrule
Nearest neighbor distance, $a_\text{NN}$ [nm] & $1.8$ & $1.8$ & $1.8$\cite{Felekidis2018} \\
LUMO acceptor, $E_\text{LUMO}^A$ [eV] & $3.8$ & $4.0$ & $4.0$\cite{Zhan2020} \\
HOMO donor, $E_\text{HOMO}^D$ [eV] & $5.2$ & $5.44$ & $5.48$\cite{Zhan2020} \\
Attempt-to-hop frequency electrons, $\nu_{0,e}$ [$\unit{s^{-1}}$] & $1 \times 10^{11}$ & $1 \times 10^{11}$ & $1.6 \times 10^{10}$\cite{Upreti2019} \\
Attempt-to-hop frequency holes, $\nu_{0,h}$ [$\unit{s^{-1}}$] & $1 \times 10^{10}$ & $1 \times 10^{11}$ & $1.6 \times 10^{9}$\cite{Upreti2019} \\
Energetic disorder electrons, $\sigma_e$ [meV] & $75$ & $70$ & $68$\cite{Upreti2019} \\
Energetic disorder holes, $\sigma_h$ [meV] & $75$ & $70$ & $89$\cite{Upreti2019} \\
Inverse exciton lifetime, $\nu_\text{ex}$ [$\unit{s^{-1}}$]	& $1 \times 10^{9}$ & $1 \times 10^{9}$ \\
Inverse CT~state lifetime, $\nu_\text{CT}$ [$\unit{s^{-1}}$] & $3 \times 10^{7}$ & $3 \times 10^{7}$ \\
Injection barrier height [eV] &	$0.2$ &	$0.2$ \\
\bottomrule
\end{tabular}
\label{tab:kmc}
\end{table}

\paragraph{Input Parameters} For the TQ1:\ce{PC71BM} system, the parameters in Table~\ref{tab:kmc} above are taken from our earlier work,\cite{Wilken2020} where we also motivate the choice of equal values for the energetic disorder for electrons and holes. In short, the choice for symmetric disorder values and attempt-to-hop frequencies that differ by an order of magnitude roughly maintains the right, experimentally determined, mobility values and ratio while keeping calculation times manageable. A similar reasoning was used for setting the parameters for the PM6:Y6 system, for which much less KMC-relevant parameters have been published previously and the uncertainty therefore is larger. As for the TQ1:\ce{PC71BM} system, we symmetrized the hopping parameters for numerical reasons. The changes in the hole parameters, which were needed to reproduce the experimental fill factor, actually lead to a reduced effect of disorder in the form of faster thermalization as compared to the values in the rightmost column. From our experience, rather substantial parameter fluctuations for different experiments performed on different batches of nominally the same material, are unfortunately not uncommon.

\paragraph{Simulation Procedures} The KMC~model was calibrated for each material system to describe the $J$--$V$ curve of a \textit{single} device thickness at a given temperature, as described in Ref.~\citenum{Wilken2020}, after which all parameters were kept constant. In the calibration step, the energy levels were adjusted to match~$\voc$, by calculating a $J$--$V$ curve using literature values for the acceptor LUMO and donor HOMO levels and subsequently adjusting one of those to make the calculated~$\voc$ equal to the measured value. Since all other parameters are kept constant, $\voc$ is linear in the effective band gap and this procedure converges in a single iteration. Other parameters like the injection barrier heights and optical constants were determined by independent experiments. The fill factor was not explicitly calibrated. The calculated dependencies on thickness and temperature are therefore model predictions and not fits. That the model captures the~($\voc$) behavior that is relevant to the present argument is not something the model was `tweaked' to do. Instead, it is a consequence of the physics that is included. KMC simulations were performed on boxes containing~$\unit[40^3]{sites}$, which for a nearest neighbor distance of~\unit[1.8]{nm} corresponds to~$\unit[72^3]{nm^3}$. Averages over multiple random configurations of the site energies were taken until the resulting error bars were sufficiently small, i.e., of the size of the symbols used in the figures. Using the described methodology, full $J$--$V$ curves can be simulated with sufficient accuracy for direct comparison with experiments. Unfortunately, both the calculation times and the numerical uncertainty tend to go up around open circuit, which precludes meaningful statements about the illumination intensity dependence of~$\voc$. Especially at sub-1-sun intensities, which is the regime at which experiments can be done without heating and degradation effects setting in, the numerical noise due to the presence of Ohmic contacts quickly overwhelms any photoinduced signals despite the use of the `buffered' contacts described above.
 
\subsection{Drift--Diffusion Model}
The drift--diffusion~(DD) model solves a set of coupled differential equations, namely the Poisson equation and the continuity equations for electrons and holes, using the one-dimensional Scharfetter–Gummel discretisation.\cite{Scharfetter1969} Charge recombination was treated in terms of a bimolecular rate equation, $R = k_2 n p$, where $k_2$ is the recombination rate constant and $n$ and $p$ the density of electrons and holes. Further details of the specific implementation can be found in Ref.~\citenum{Felekidis2018}. To take into account the (Gaussian) disorder and to make the DD~model as comparable as possible to the KMC model, three measures were implemented:
\begin{enumerate} 
\item The generalized Einstein equation\cite{Roichman2002} was used for the relation between the diffusion coefficient and the mobility.
\item As boundary conditions at the contacts, the interfacial charge density was used, which was explicitly calculated as the integral over the Gaussian DOS multiplied with the Fermi–Dirac distribution, see Eq.~(\ref{eq:Fermi}). That means that the same boundary condition is used as for the `buffered' contacts in the KMC~model.\cite{vanderHolst2009,Paasch2010} For non-Ohmic contacts, defined as having an electric field~$F_C$ at the semiconductor/metal interface that directs a drift current away from the metal, the injection barrier is lowered by the image potential as:\cite{vanderHolst2009}
\begin{equation}
\Delta' = \Delta - 	q \sqrt{\frac{q|F_C|}{4\pi\varepsilon_0\varepsilon_r}}
\end{equation}
For $F_C$ directing a drift current towards the metal~(Ohmic contacts) the full injection barrier~$\Delta$ is used.
\item The mobility functional for the extended Gaussian disorder model by Pasveer~et~al.\cite{Pasveer2005} was applied to translate the same set of hopping parameters~($a_\text{NN}$, $\nu_0$, $\sigma$) as used in the KMC~model into a quasi-equilibrium mobility. It was checked that the KMC~model reproduces the mobility values from Pasveer~et~al. Hence, for thermalized charge carriers, KMC and DD use the same mobility values; evidently, since in DD all charge carriers are assumed to be thermalized, the in KMC naturally included effect that nonthermalized charge are more mobile is absent in DD. 
\end{enumerate} 

All relevant input parameters for the DD~model are given in Table~\ref{tab:dd}. Note that all parameters are equal to the corresponding numbers used in the KMC model~(Table~\ref{tab:kmc}) with the exception of the recombination rate constant that is not an independent parameter in KMC and that was taken from literature.\cite{Kniepert2019} The effective energy gap is calculated as the difference between the donor HOMO energy and the acceptor LUMO energy.

\begin{table}[t]
\caption{Overview of the parameters used for the drift--diffusion simulation of TQ1:\ce{PC71BM} and PM6:Y6 solar cells.}
\begin{tabular}{lcc}
\toprule
Parameter [unit] & \multicolumn{2}{c}{Value}\\
 \cmidrule(lr){2-3}
 & TQ1:\ce{PC71BM} & PM6:Y6 \\
\midrule
Effective energy gap [eV] & $1.4$ & $1.44$ \\ 
Nearest neighbor distance, $a_\text{NN}$ [nm] & $1.8$ & $1.8$  \\
Attempt-to-hop frequency electrons, $\nu_{0,e}$ [$\unit{s^{-1}}$] & $1 \times 10^{11}$ & $1 \times 10^{11}$  \\
Attempt-to-hop frequency holes, $\nu_{0,h}$ [$\unit{s^{-1}}$] & $1 \times 10^{10}$ & $1 \times 10^{11}$  \\
Energetic disorder [meV] & $75$ & $70$ \\
Recombination rate constant, $k_2$ [$\unit{m^3s^{-1}}$] &  $2 \times 10^{-17}$ &  $2 \times 10^{-17}$ \\
Injection barrier height [eV] &	$0.2$ &	$0.2$ \\
\bottomrule
\end{tabular}
\label{tab:dd}
\end{table}

\subsection{Optical Modeling}
\label{sec:tmm}
\begin{figure}
\includegraphics[width=0.66\textwidth]{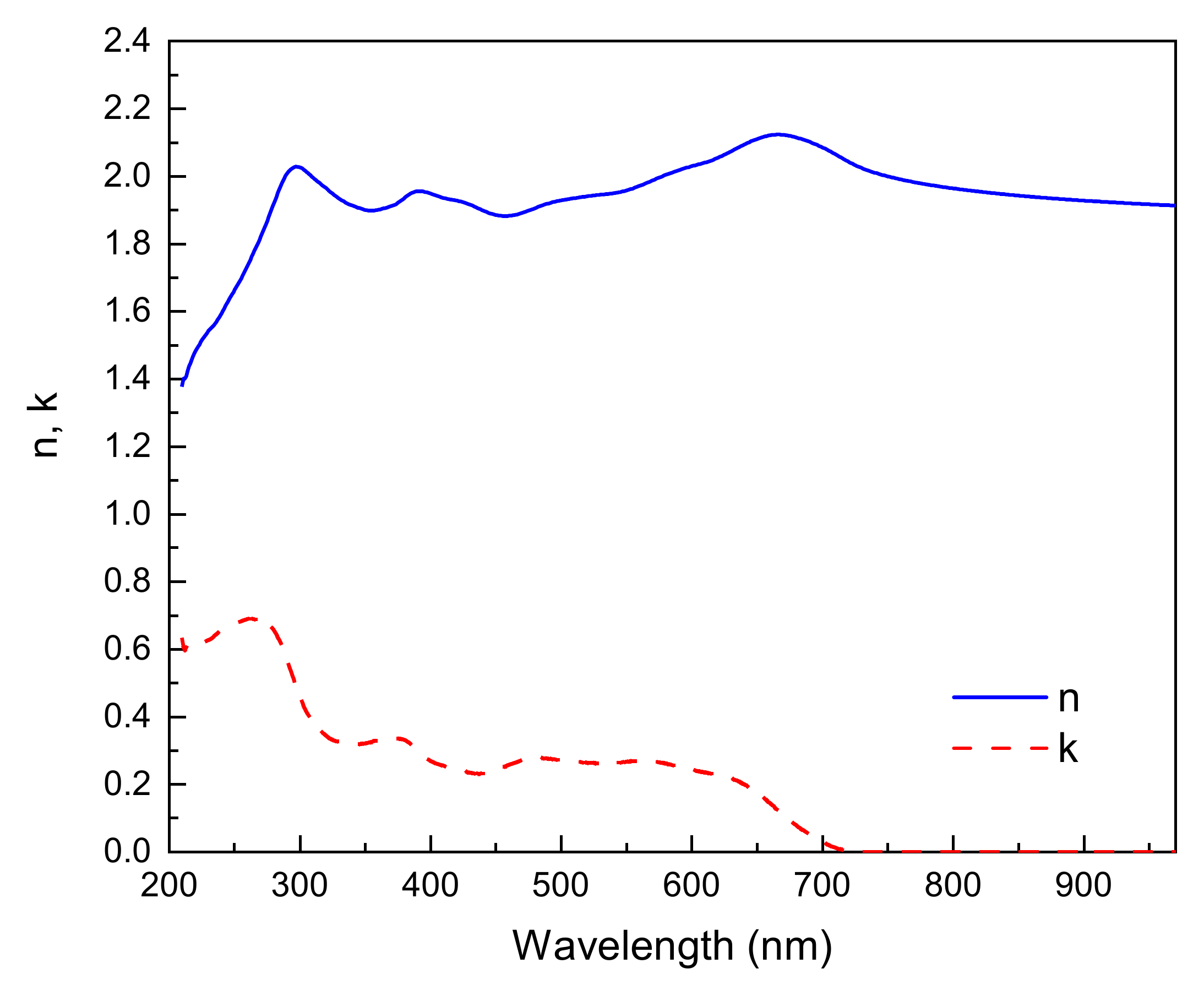}
\caption{Optical constants for TQ1:\ce{PC71BM} as used in the transfer matrix modeling.}
\label{fig:figS2}
\end{figure}

To take into account the spatial profile of the optical generation rate and its variation with thickness, the spatial distribution of photocreated singlet excitons was calculated for the TQ1:\ce{PC71BM} system using the transfer-matrix approach as described previously.\cite{Burkhard2010} The used optical constants were determined by spectroscopic ellipsometry and are shown in Figure~\ref{fig:figS2}. Since no systematic thickness-dependent study was performed for the PM6:Y6 system, optical modeling was not needed, and a constant generation rate was used.

\pagebreak
\section{Reciprocity Analysis of KMC~Simulations}
In our reciprocity analysis we considered an extended version of Eqs.~1 and 2 in the main text,
\begin{equation}
\voc = \frac{kT}{q} \ln\left(\frac{J_\text{gen}}{J_0} + 1\right) + \frac{kT}{q} \ln(\text{EQE}_\text{EL}) + \frac{kT}{q} \ln \left(\frac{F_\text{coll}}{F_\text{inj}}\right)
\label{eq:reciproc1}
\end{equation}
Here, the first term on the right-hand side is the radiative limit for~$\voc$. The second term accounts for losses due to nonradiative recombination. The third term accounts for differences in collection efficiency~(during photovoltaic operation under illumination) and injection efficiency~(during operation as light emitting device in the dark), as derived by Kirchartz~et~al.\cite{Kirchartz2016} The value for~$\text{EQE}_\text{EL}$ can be determined from KMC~simulations by running the same device as light emitting diode, i.e., in the dark under a forward bias corresponding to~$\voc$, and dividing the integrated recombination current by the total injection current,\cite{Felekidis2020}
\begin{equation}
\text{EQE}_\text{EL} = \frac{J_\text{EL}}{J_\text{inj}}
\end{equation}
For the used parameters and morphology, a value of~$\text{EQE}_\text{EL} = 0.14$ was found for TQ1:\ce{PC71BM}. Likewise, the values for~$F_\text{inj}$ and~$F_\text{coll}$ can be obtained from KMC~simulations at~$V = \voc$ using
\begin{equation}
F_\text{inj} = \frac{1}{d} \int_0^d \frac{n(x,V) p(x,V) - n_i^2}{n_i^2(\exp(qV/kT)-1)} \diffd x
\end{equation}
where $n$ and $p$ are the electron and hole densities, $d$ the active-layer thickness and $n_i$ the intrinsic (thermal) charge density corresponding to the effective energy gap, $n_i^2 = N_0^2 \exp(-E_\text{gap}^\text{eff}/kT)$ with the site density~$N_0 = a_\text{NN}^{-3}$, and
\begin{equation}
F_\text{coll} = \frac{1}{d} \int_0^d f_c(x,V) \diffd x
\end{equation}
Where, $f_c(x,V)$ is the collection probability of charges photogenerated at a position~$x$ in the device. The spatially averaged values for electrons and holes are the same, and equal to the ratio of the photocurrent~(light minus dark current) and the maximum generated photocurrent~$J_\text{gen}$ that is either determined by optical modeling~(see Section~\ref{sec:tmm} above) or as~$J_\text{gen} = qG_\text{av}d$ with  $G_\text{av}$ the average exciton generation rate.\cite{Kirchartz2016} Hence,
\begin{equation}
F_\text{coll} = \frac{J_\text{photo}}{J_\text{gen}}
\end{equation}
For the TQ1:\ce{PC71BM} system, for which this analysis is performed, $F_\text{inj}$ and~$F_\text{coll}$ were found to be equal, with a value~$F_\text{inj} = F_\text{coll} = 0.32$, meaning that the last term in Eq.~(\ref{eq:reciproc1}) actually becomes zero.

Finally, the reverse dark saturation current can be obtained by integrating Eq.~(2) of the main text as
\begin{equation}
J_0 = q \int \text{EQE}_\text{PV}(E)\phi_\text{BB}(E) \diffd E
\label{eq:J0}
\end{equation}
for which we use
\begin{equation}
\text{EQE}_\text{PV}(E) = \text{IQE}_\text{PV}(E) \phi_\text{abs}(E)
\label{eq:EQEpv}
\end{equation}
with $\text{IQE}_\text{PV}$ the internal quantum efficiency for photovoltaic operation that is set to unity unless stated otherwise\cite{Felekidis2020} and~$\phi_\text{abs}$ the absorption spectrum. Due to the steepness of the black body spectrum~$\phi_\text{BB}$,
\begin{equation}
\phi_\text{BB}(E) = \frac{2\pi E^2}{h^3c^2}\frac{1}{\exp(E/kT)-1}\quad[\unit{m^{-2}s^{-1}J^{-1}}],
\label{eq:blackbody}
\end{equation}
only the energetically lowest parts of the CT and S1 contributions to~$\phi_\text{abs}$ are important. We write~$\phi_\text{abs}$ as
\begin{equation}
\phi_\text{abs}(E) = a \phi_\text{CT}(E) + b \phi_\text{S1}(E),
\label{eq:phiabs}
\end{equation}
where the (TQ1:\ce{PC71BM}) CT and (TQ1) singlet absorption spectra are calculated as convolutions of the relevant HOMO and LUMO levels as described before;\cite{Felekidis2020} their central energies are corrected for the Coulomb binding energies of the S1~(\unit[0.5]{eV}) and CT~(\unit[0.22]{eV}) states. The weight factors~$a$ and~$b$ are estimated as follows. Since TQ1 is a strong absorber, we take~$b = 1$ at the absorption maximum, i.e., all photons impinging on the sample with the energy of the absorption maximum get absorbed. The factor~$a$ can then be estimated from
\begin{equation}
\frac{a}{b} = \frac{\nu_\text{CT}}{\nu_\text{S1}}\frac{n_{s,\text{CT}}}{n_{s,\text{S1}}}
\label{eq:reciproc2}
\end{equation}
where $\nu_\text{S1} = \unit[1 \times 10^9]{s^{-1}}$ and $\nu_\text{CT} = \unit[3 \times 10^7]{s^{-1}}$ are the S1 and CT~recombination rates as used in the model~(see Table~\ref{tab:kmc}); these values have been calibrated to recombination transients.\cite{Wilken2020} Equation~(\ref{eq:reciproc2}) makes the reasonable assumption that CT and S1~recombination are competing against the same or at least similar loss channels, such that their relative lifetimes reflect their relative oscillator strengths. The second term on the right-hand side of Eq.~(\ref{eq:reciproc2}) accounts for the fact that the number of absorption sites in the simulation box, $n_s$, is different for S1 and CT~absorption. For the simplified morphology used here, a~$10 \times 10$~unit cell with $7 \times 7$~inclusions of aggregated \ce{PC71BM}, the lowest CT~states are found at the interface between the mixed matrix and the \ce{PC71BM} inclusions, giving rise to a ratio~$n_{s,\text{CT}}/n_{s,\text{S1}} = 28/51 \approx 0.55$.

\begin{figure}
\includegraphics[width=0.66\textwidth]{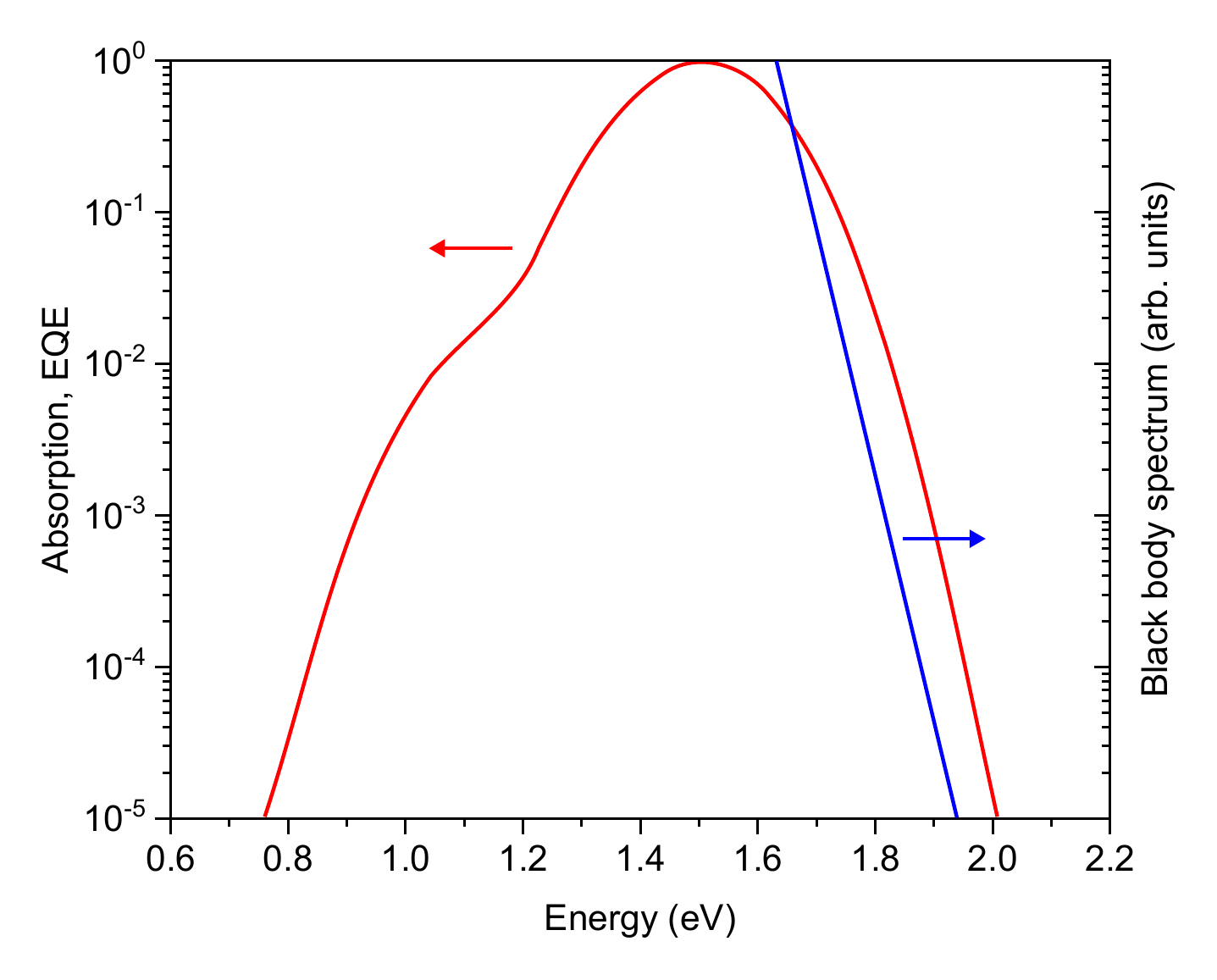}
\caption{Absorption spectrum calculated from the parameters in Table~\ref{tab:kmc} and Eqs.~(\ref{eq:phiabs}) and~(\ref{eq:reciproc2}). The CT~contribution is visible as a hump on the low-energy side of the spectrum. The main peak is the S1~contribution. The black body spectrum for~$T = \unit[300]{K}$, Eq.~(\ref{eq:blackbody}), is shown in blue. The donor LUMO energy, which is not needed in the KMC~simulations, is taken~$E_\text{LUMO}^D = \unit[3.2]{eV}$. Since only the lowest part of the absorption spectrum is relevant for the calculation of $J_0$, Eqs.~(\ref{eq:J0}) and~(\ref{eq:EQEpv}), the more or less flat continuum that is found in experiments at energies beyond the S1~peak is not shown.}
\label{fig:figS3}
\end{figure}

The parameters listed in Table~\ref{tab:kmc} give rise to the absorption spectrum in Figure~\ref{fig:figS3}, where also the black body spectrum is shown. Despite the simplifications made, the ratio of the S1 and CT~absorption peaks of $\sim$0.017 that follows from Eq.~(\ref{eq:reciproc2}) is consistent with the experimentally observed value that falls in the range of~0.01 to 0.033, depending on whether on considers only the 0--0~transition, as done here, or the full vibronic progression.\cite{Felekidis2020} This suggests that the assumptions made are reasonable from a physical perspective. From Eqs.~(\ref{eq:reciproc1})--(\ref{eq:reciproc2}) we then obtain an equilibrium value of~$\voc \approx \unit[0.69]{V}$.

\begin{figure}
\includegraphics[width=0.66\textwidth]{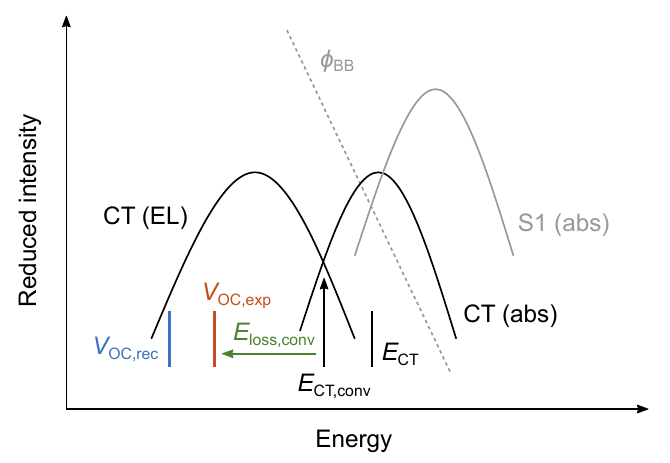}
\caption{Sketch of the energetics and optical spectra. The vertical lines show the relative positions of characteristic energies with respect to the CT electroluminescence~(EL) and absorption~(abs) spectra and the red tail of the singlet~(S1) absorption spectrum. The measured~$V_\text{OC,exp}$ is conventionally interpreted in terms of a loss energy w.r.t.~the intersection of the CT~EL and absorption spectra, labeled~$E_\text{CT,conv}$, that is referred to as the (relaxed) CT~state. In our earlier work,\cite{Puttisong2018} we have shown that the actual relaxed CT~energy lies at much lower energies~(in fact, slightly below~$V_\text{OC,exp}$ for the TQ1:\ce{PC71BM} system, not shown), and is therefore not a relevant point of reference. The central CT~energy, labeled~$E_\text{CT}$, lies above~$E_\text{CT,conv}$ but is not directly experimentally accessible. Assuming electron and hole populations are in thermodynamic equilibrium within their respective bands, as in the reciprocity analysis performed above, leads to a prediction for~$\voc$ that underestimates the actual value by~$\sim$\unit[0.2]{eV}, c.f.~the blue line labeled~$V_\text{OC,rec}$. Using the same starting parameters in a KMC~model, which makes no upfront assumptions about charge populations being in equilibrium, does reproduce the measured open-circuit value, as shown in this work.}
\label{fig:figS4}
\end{figure}

\pagebreak
\section{Transient Energetics of Photogenerated Charges}
\begin{figure}
\includegraphics[width=0.7\textwidth]{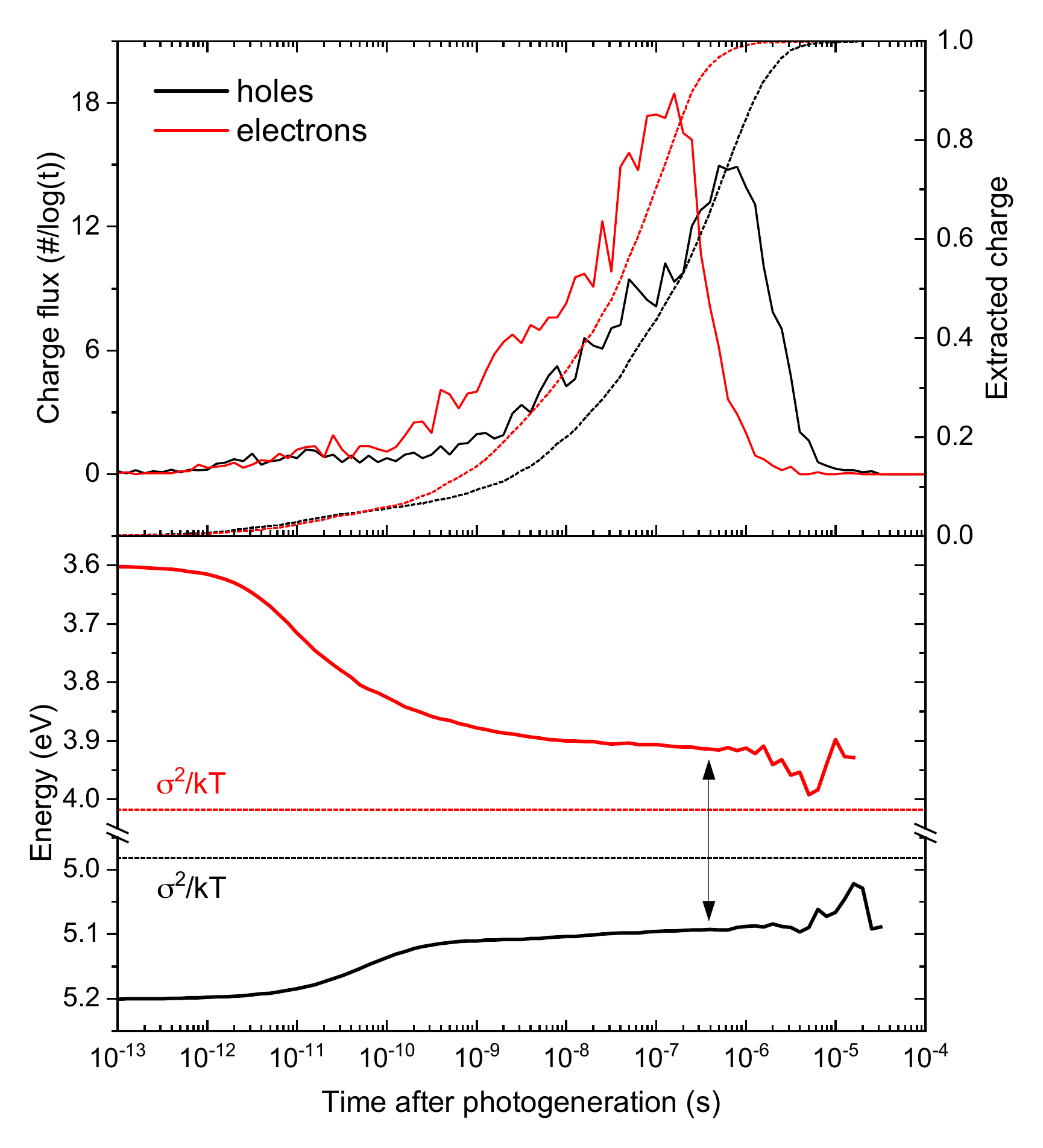}
\caption{Energy loss after photoexcitation under open-circuit conditions for TQ1:\ce{PC71BM}. The top panel shows the extraction time distribution of photogenerated electrons~(red) and holes~(black) as solid lines, the corresponding integrated fraction of extracted charge is shown as dotted lines. The bottom panel shows the corresponding thermalization of photogenerated charges, the dotted horizontal lines indicate the equilibrium energies that lie~$\sigma^2/kT$ below~(above) the LUMO~(HOMO) energy. Note that the electrons are generated in the mixed TQ1:\ce{PC71BM} phase and loose an additional~\unit[0.2]{eV} upon transfer to the pure \ce{PC71BM} phase, which explains their larger apparent energy loss as compared to holes that remain in the mixed phase.}
\label{fig:figS5}
\end{figure}

From the KMC calculations, the energy distribution of photogenerated charge carriers can be traced as a function of time after photogeneration, i.e., the electron and hole energies are followed as a function of time after the exciton from which they originate was generated. This calculation is done under steady state conditions and is thus relevant to device operation. Similar results as shown in Figure~\ref{fig:figS5} below were previously shown in Refs.~\citenum{Wilken2020,Melianas2015,Melianas2014,Felekidis2018b} The results show that even under short-circuit conditions, charge thermalization does not complete before the charge carriers have been extracted. This can be seen from the fact that the mean electron and hole energies~(solid lines, lower panel) have not reached the equilibrium energy~(dashed lines) at the time where essentially all charges have left the device by either extraction or recombination. For the used parameters, this is around~$t = \unit[10^{-6}]{s}$ and~$t = \unit[10^{-5}]{s}$ for electrons and holes, respectively, see the upper panel.

At higher charge carrier densities, thermalization is no longer to the equilibrium energy but to the (quasi-)Fermi level, which would be an alternative interpretation of the data in Figure~\ref{fig:figS5}. However, to further highlight that thermalization is incomplete and does not stop at a (quasi-)Fermi level, which would be an indication of local equilibrium, we note that the difference between the mean electron and hole energies, indicated by the arrow in the bottom panel, saturates at around 1.15--\unit[1.20]{eV}, which is~$\sim$\unit[0.3]{eV} above the measured and calculated~(by KMC) open-circuit voltage. The possibility that thermalization in Figure~\ref{fig:figS5} might stop due to charges reaching a quasi-Fermi level can also be ruled out from the fact that this would require the quasi-Fermi levels to lie~$\sim$\unit[0.1]{eV}, i.e., less than~$2\sigma$, below the center of the band, which in turn would require charge densities that are orders of magnitude larger than typically observed in organic solar cells under open-circuit conditions. This corroborates the picture sketched in Fig.~3 of the main text, in which, at~$V = \voc$, a nonthermalized photocurrent is counteracted by an injection current coming from the (thermalized) contacts.

One could expect signatures of the nonthermalized populations of the photogenerated electrons and holes in plots of calculated densities occupied states vs.~energy. Unfortunately, even after several days of computation at a single voltage point, the statistics are not enough to discern any meaningful high energy tails to, or to determine differences in effective temperature from. The reason is that the statistically fluctuating charge carrier population of charges diffusing in from the contacts overwhelms any fast moving photogenerated charges.

Interestingly, it is possible to get further confirmation for the existence of nonequilibrium currents from the energy resolved current distributions, see Figure~\ref{fig:figS6} below. We attribute the oscillating shape of the current densities in the dark to the way these plots are extracted from the KMC~calculations, as further explained in the caption. More interesting is the difference between the current distributions under illumination and in the dark~(blue lines), which especially in the case of short-circuit conditions shows a clear negative~(since $\jsc < 0$) peak at~$\sim$\unit[5.15]{eV}, i.e., near to the center of the DOS, that lies well above both the maximum of the density of occupied states and the oscillating background current distribution. Interestingly, the center of this high-energy peak lies at roughly the same energy as the mean energy of the relaxing hole distribution~(black line in the lower panel of Figure~\ref{fig:figS5}), which is consistent with the notion that the photocurrent is predominantly carried by nonthermalized holes. Under open-circuit conditions~(dashed lines), there is no clear peak at higher energies, but instead there is an evident negative tail in the same region~($\sim$\unit[5.1]{eV} and upwards). Note that the total area under the curve is zero at open-circuit conditions under illumination.

On basis of the results in Figures~\ref{fig:figS5} and~\ref{fig:figS6} and the schematic Figure~\ref{fig:figS4}, one would expect a different position for the emission maximum between~EL, which originates from `equilibrated' charges coming from the contacts and PL, which would come from only partially thermalized photogenerated charges. This topic was addressed before in Ref.~\citenum{Felekidis2020} (where also experimental data are shown in Fig.~S9). In short, KMC~simulations do show a minor effect of the expected sign. However, the magnitude is very small due to `equilibrium' EL also coming from a nonthermal subset of sites,\cite{Melianas2019} and is much smaller than the experimentally observed difference, which we attribute to sample inhomogeneity.\cite{Felekidis2020}

The area under the energy-resolved current curves reflects the corresponding net hole current density. All curves have been averaged over the full device thickness to reach sufficient statistics. We believe the oscillating shape in absence of illumination to be due to a combination of two factors. First, they reflect the presence of an Ohmic contact in which a positive diffusion current is preferentially injected into low-lying empty states and is compensated by a negative drift current that flows closer to the transport energy. Second, since the curves had to be averaged over the full device, spatial variations in charge and current distributions in combination with band bending may cause additional features. It is important that the implemented contacts do not act as charge pumps, i.e. in absence of illumination, $J$--$V$~curves go through the origin.

\pagebreak

\pagebreak
\begin{figure}
\includegraphics[width=0.66\textwidth]{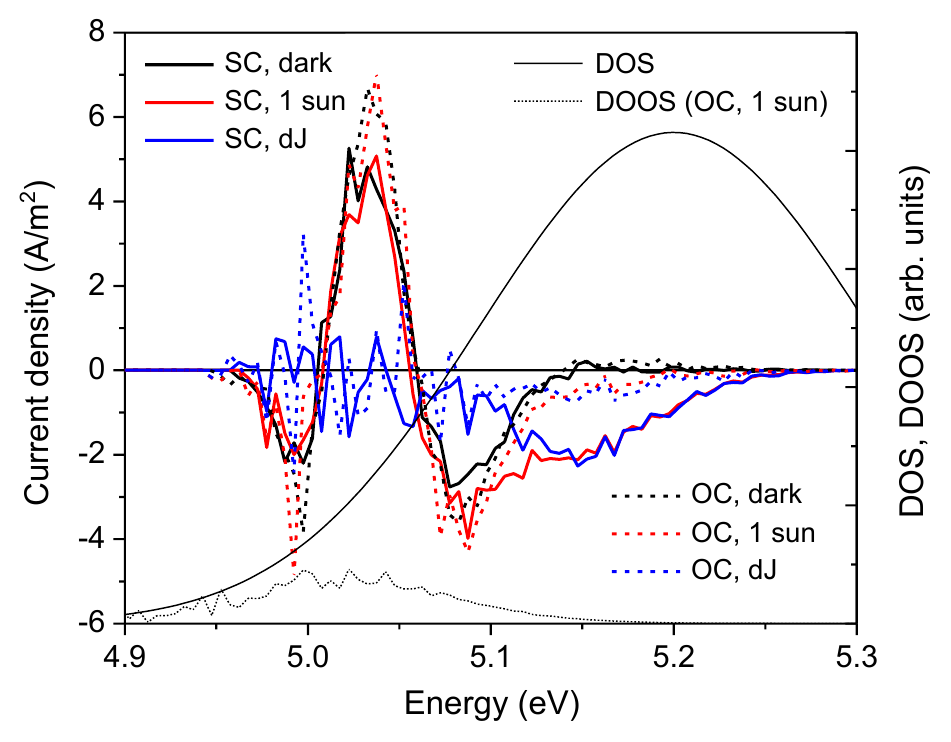}
\caption{Energy-resolved hole currents for TQ1:\ce{PC71BM} w.r.t.~the HOMO DOS. The solid thick lines show KMC-calculated current densities under short circuit~(SC, solid) and open circuit~(OC, dashed) conditions, the blue lines are the differences between the currents in the dark~(black) and under illumination~(red). The thin solid and dashed lines indicate the density of states~(DOS) and density of occupied states~(DOOS) for reference.}
\label{fig:figS6}
\end{figure}

\pagebreak
\section{Current--Voltage Curves at Large Reverse Bias}
\begin{figure}
\includegraphics[width=\textwidth]{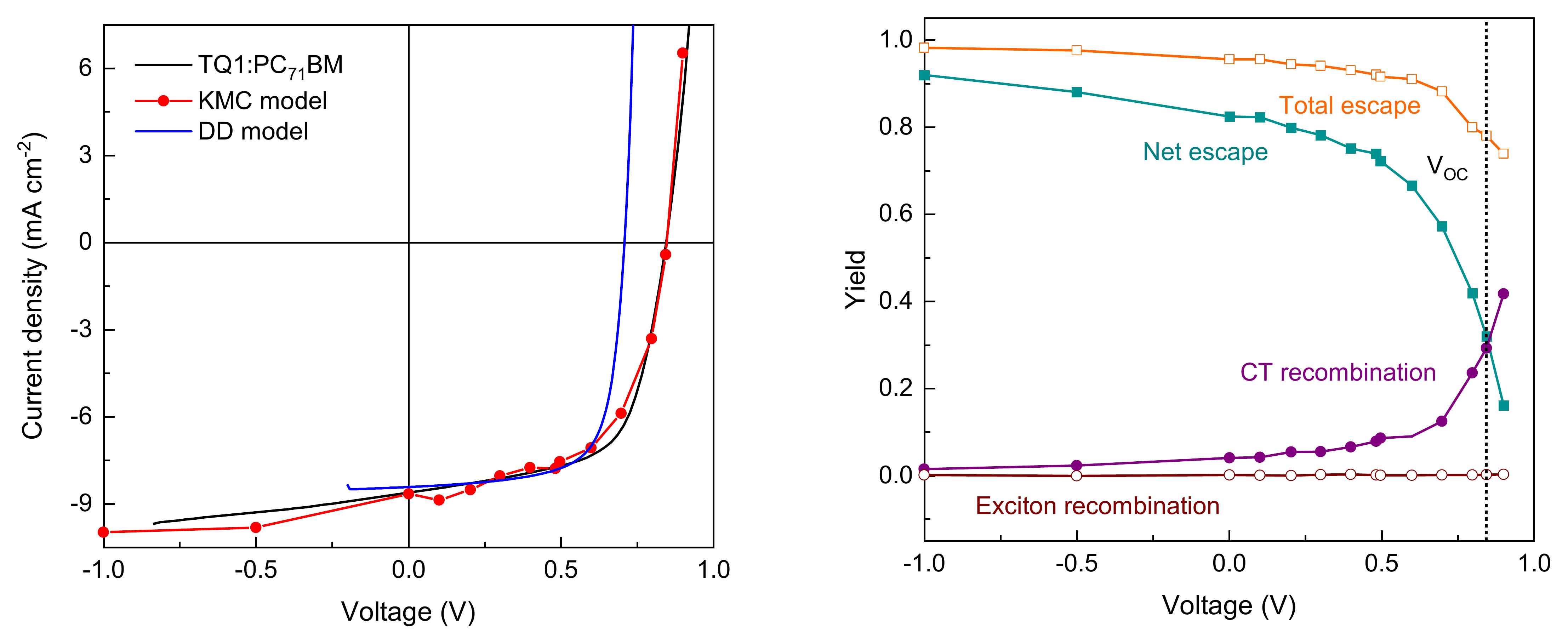}
\caption{Measured versus modelled $J$--$V$ curves and loss analysis. Same data as in Fig.~1 of the main text for a wider revers bias range. (a)~The black line represents measured $J$--$V$~characteristics of a 70-nm thick TQ1:\ce{PC71BM} solar cell under illumination. Only the KMC~model~(red symbols) reproduces the gradual increase in current density towards larger reverse bias, as observed in the experiment. The DD~model~(blue line) predicts a constant current beyond~$V = 0$. The KMC and DD models use a single, consistent set of parameters. (b)~Corresponding extraction and loss yields from KMC. Total and net escape yields are defined as~$y_\text{total} = (J_{n,\text{an}} + J_{n,\text{cat}} + J_{p,\text{an}} + J_{p,\text{cat}})/J_\text{abs}$ and $y_\text{net} = (-J_{n,\text{an}} + J_{n,\text{cat}} +J_{p,\text{an}} - J_{p,\text{cat}})/J_\text{abs}$, where $J_{(n/p),(\text{an}/\text{cat})}$ is the current density of photogenerated electrons/holes extracted via the anode/cathode and $J_\text{abs}$ is the current density corresponding to light absorption. The curves labelled exciton and CT~recombination show the relative current densities associated with exciton and CT~recombination, i.e., the fraction of photogenerated charges that undergo these processes.}
\label{fig:figS7}
\end{figure}

\pagebreak
\section{Temperature Dependent Drift--Diffusion Simulations}
\begin{figure}
\includegraphics[width=0.66\textwidth]{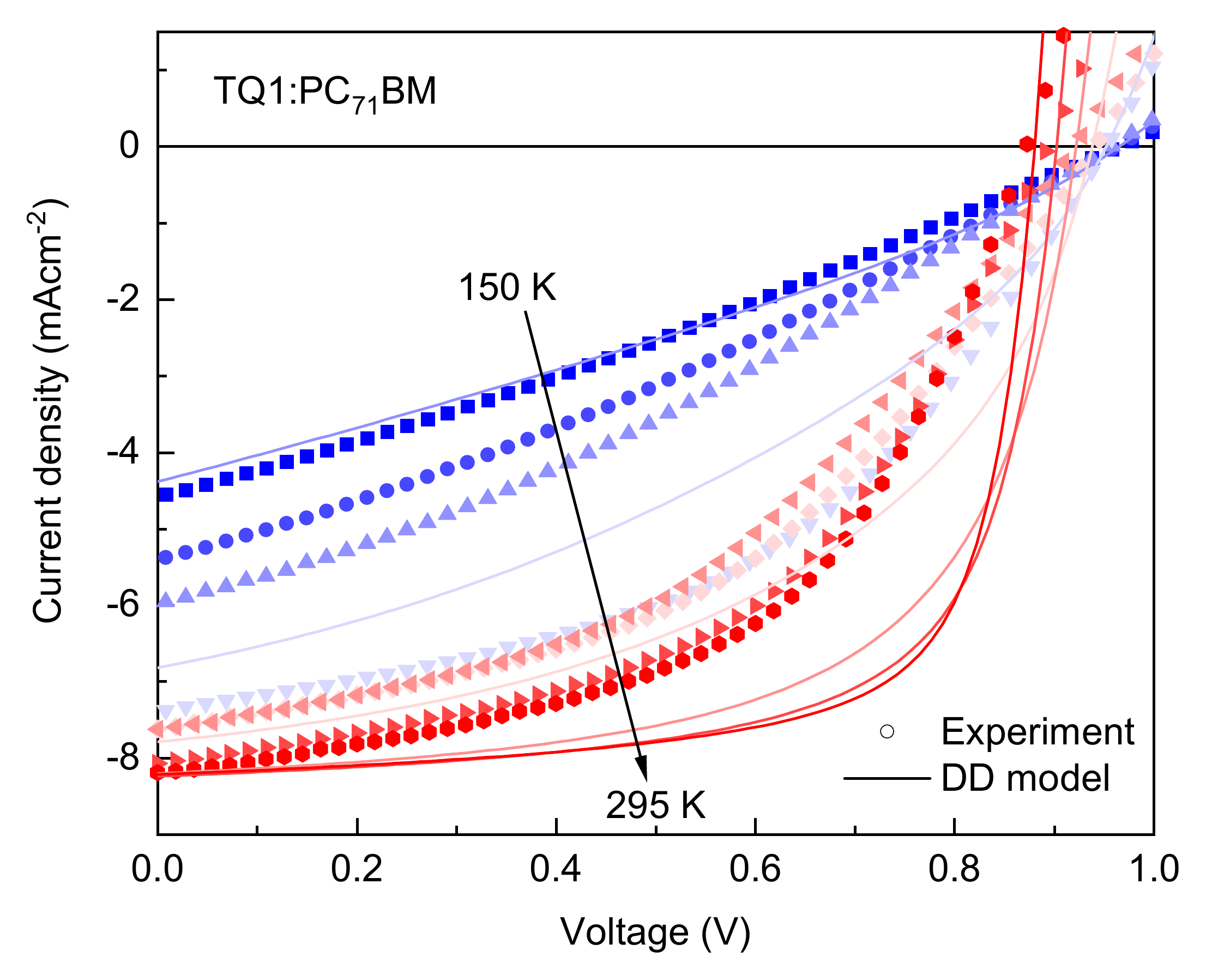}
\caption{Temperature dependent $J$--$V$ characteristics for a 75-nm TQ1:\ce{PC71BM} device~(symbols, same data as in Fig.~4a of the main text) and DD simulations~(lines) with the parameters from Table~\ref{tab:dd}. The simulations have been shifted to match~$\voc$.}
\label{fig:figS8}
\end{figure}

\pagebreak
\begin{figure}
\includegraphics[width=0.66\textwidth]{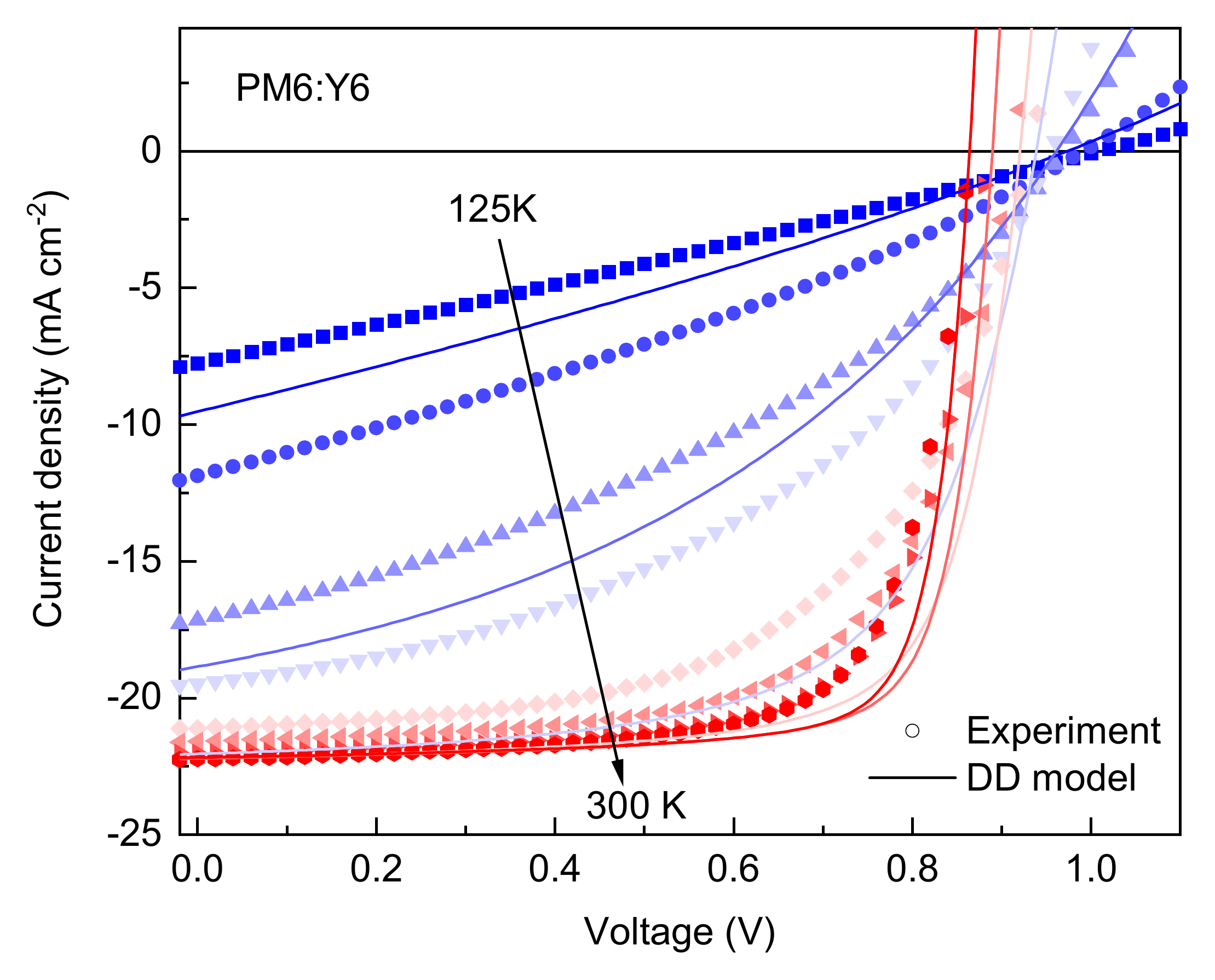}
\caption{Temperature dependent $J$--$V$ characteristics for a 115-nm PM6:Y6 device~(symbols, same data as in Fig.~4c of the main text) and DD simulations~(lines) with the parameters from Table~\ref{tab:dd} that have been shifted to match~$\voc$. Note the failure of the latter to describe the measured evolution of the fill factor with temperature.}
\label{fig:figS09}
\end{figure}

\pagebreak
\section{Temperature Dependence for a Thick TQ1:\ce{PC71BM} Device}

\begin{figure}
\includegraphics[width=\textwidth]{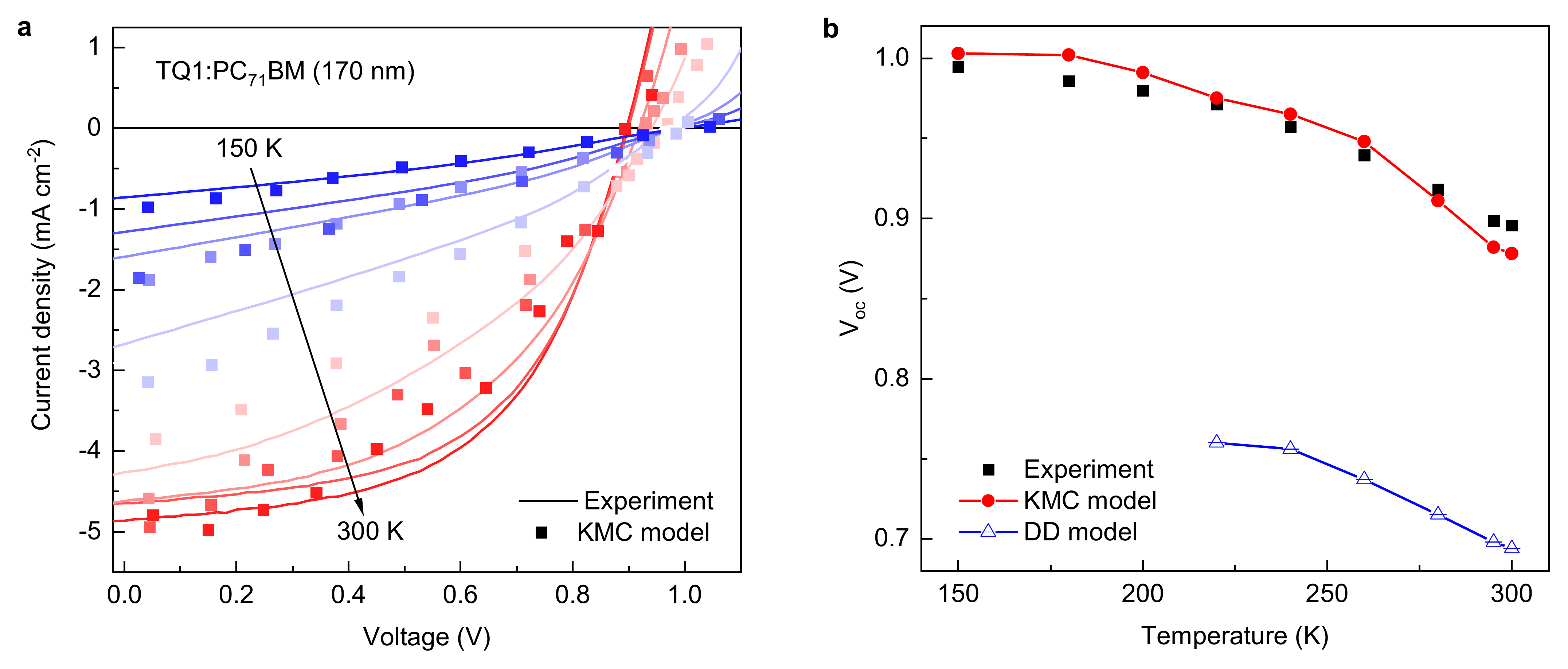}
\caption{(a)~Measured and KMC-modeled temperature dependent $J$--$V$ characteristics for a 170-nm thick TQ1:\ce{PC71BM} solar cell. (b)~Corresponding~$\voc$ at different temperature compared with the DD~model. Due to numerical instabilities, the DD~model did not converge below~$\sim$\unit[230]{K}. Note in this figure that the (forward) bias dependence of the $J$--$V$~curves around room temperature is suppressed as compared to the thin device shown in Figs.~1 and~4 of the main text and in Figure~\ref{fig:figS7} above. This is consistent with the interpretation in terms of a diffusion loss due to highly diffusive `hot' charge carriers: for thicker devices, a smaller fraction of carriers is generated within the diffusion distance from the contact. Consequently, the agreement of DD~simulations with experiments at large reverse bias improves for thicker devices around room temperature. }
\label{fig:figS10}
\end{figure}

\pagebreak
\section{Role of Energetic Disorder}
\begin{figure}
\includegraphics[width=0.66\textwidth]{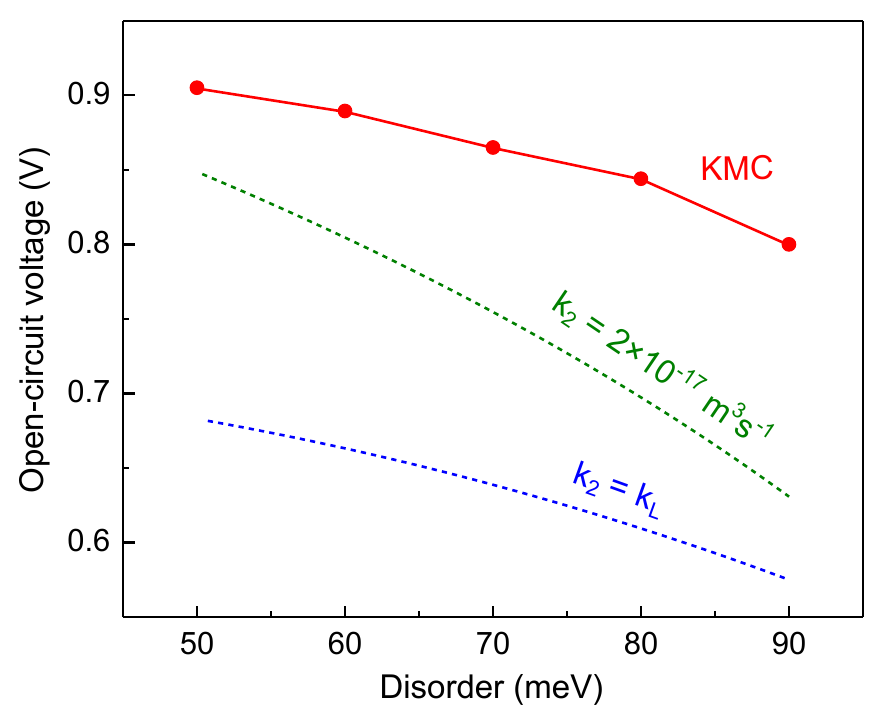}
\caption{Role of disorder on the difference between equilibrium and nonequilibrium models. Symbols indicate the results of KMC~simulations using the parameters for the PM6:Y6 system~(Table~\ref{tab:kmc}) but with varying energetic disorder~$\sigma$. The lines are the predictions from the equilibrium model by Blakesley and Neher\cite{Blakesley2011} for the same parameters, using either the constant low bimolecular recombination rate from Table~\ref{tab:dd}~(green dashed line) or the Langevin value~(blue dashed line).}
\label{fig:figS11}
\end{figure}

Comparing the predictions from the KMC~model with those from equilibrium models is complicated by the fact that the bimolecular recombination rate is emergent in KMC but must be explicitly parametrized in equilibrium models as those by Blakesley and Neher\cite{Blakesley2011} or drift--diffusion; typically a reduced Langevin rate is used. Since the actual reduction value depends critically on essentially all KMC~parameters, including disorder, the equilibrium~$\voc$ prediction becomes somewhat arbitrary. Therefore, Figure~\ref{fig:figS11} shows two limiting cases: a constant low bimolecular recombination rate~($k_2 = \unit[2 \times 10^{-17}]{m^3s^{-1}}$ from Table~\ref{tab:dd}, green dashed line) and the Langevin value ($k_L = q/\varepsilon_0\varepsilon_r(\mu_e + \mu_h)$ with mobilities~$\mu_{e,h}$ that correspond to the used disorder and hopping parameters, blue dashed line). Since the actual bimolecular recombination rate is significantly reduced relative to the Langevin value due to the morphology and re-splitting of interfacial CT~states, the former curve is the more relevant one for the current system. In line with the finding in the main text that the difference between KMC and DD is larger for the more disordered TQ1:\ce{PC71BM} system than for the less disordered PM6:Y6 system~(\unit[0.17]{V} vs.~\unit[0.13]{V}, see Fig.~4 in the main text), the difference with the nonequilibrium KMC~model becomes smaller for lower disorder values.

\pagebreak
\section{Yields for Drift--Diffusion}

\begin{figure}
\includegraphics[width=0.66\textwidth]{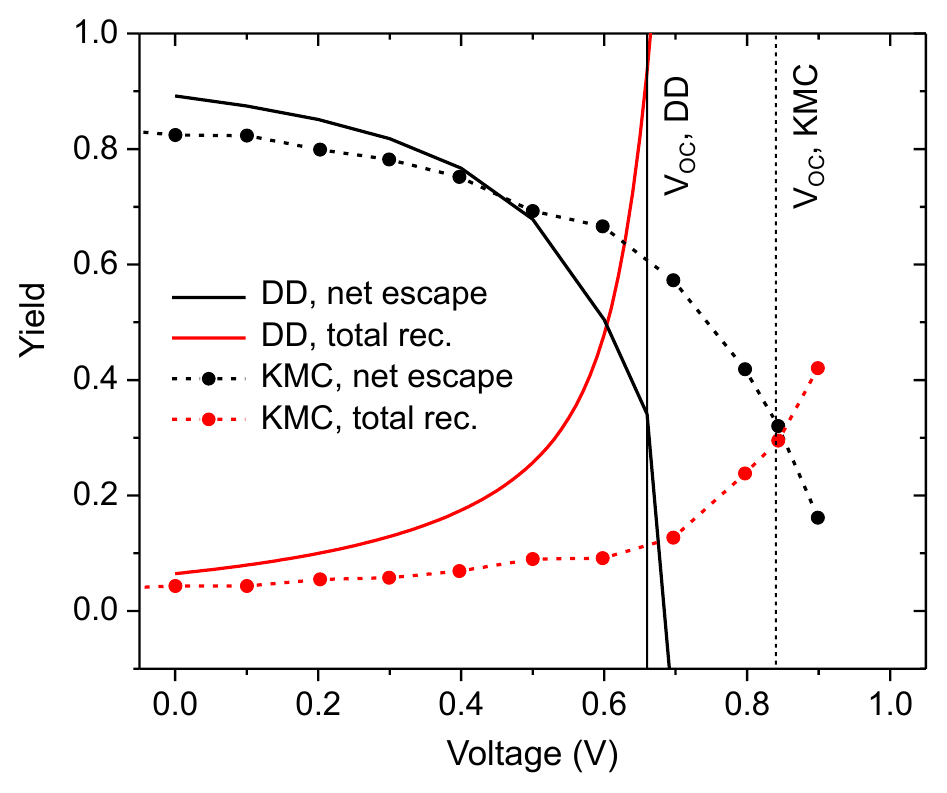}
\caption{Comparison between recombination and escape yields from KMC and DD~models. Since the DD formalism does not allow to label charges as being photogenerated or injected from the contacts but instead works with a single distribution per type of charge carrier~(electrons or holes), one cannot produce an equivalent graph to Fig.~1b in the main text. Likewise, there is only a single recombination rate in drift--diffusion. However, one can define a total recombination yield as the recombination current normalized to the maximum photocurrent~$J_\text{gen} = q G d$ and compare this to the total (CT + exciton recombination = geminate + nongeminate) recombination from KMC. Likewise, one can use the methodology from Kirchartz~et~al.\cite{Kirchartz2016} to determine a net escape yield from DD.}
\label{fig:figS12}
\end{figure}

Figure~\ref{fig:figS12} shows the yields for charge extraction and recombination from the different models. Comparing the KMC~(symbols) and DD~(solid lines) simulations, one notices that the bias dependence is stronger in the latter case, especially for the recombination. For both KMC and DD, the upswing in recombination~(red lines) around~$\voc$ is strongly affected by recombination of charges that are injected from the contacts. As argued in the main text, the voltage dependence of the escape curve~(black lines) for KMC largely reflects increasing diffusion losses. The limited information that can be extracted from the DD~simulation does not allow a similar assignment, but the strong and field dependent recombination suggests that, especially around~$\voc$, the losses are mostly due to recombination. Note also that slightly beyond~$\voc$ the escape yield for DD becomes negative, indicating that all photogenerated charges recombine, which is not observed in KMC. As such, the fact that the escape yields for KMC and DD are roughly the same at exactly open-circuit conditions should not be overinterpreted as a sign of equivalence of the two models.

\pagebreak
\bibliography{supplement}